\documentclass[10pt]{iopart}
\usepackage{iopams} 

\expandafter\let\csname equation*\endcsname\relax
\expandafter\let\csname endequation*\endcsname\relax
\usepackage{amsmath}

\usepackage{graphicx}
\usepackage{bm}
\usepackage{etoolbox}
\usepackage{physics}
\usepackage{stackengine}
\usepackage{mathtools}
\usepackage{lipsum}
\usepackage{breqn}
\usepackage{hyperref}
\usepackage{nicefrac}
\usepackage{dblfloatfix}
\usepackage{afterpage}
\usepackage{amsmath,amsfonts}
\usepackage{siunitx}
\usepackage{subfig}
\usepackage{cuted}
\usepackage{cleveref}
\usepackage{cite}
\usepackage{float}
\usepackage{siunitx}
\usepackage{graphicx}

\newcommand{\rulesep}{\unskip\ \hrule\ }

\def\XXint#1#2#3{{\setbox0=\hbox{$#1{#2#3}{\int}$ }
\vcenter{\hbox{$#2#3$ }}\kern-.6\wd0}}

\newcommand{\Alfven}{Alfv\'{e}n }
\newcommand{\Alfvenic}{Alfv\'{e}nic }

\newcommand{\Poincare}{Poincar\'{e} }

\hypersetup{
    colorlinks=true,
    linkcolor=blue,     
    urlcolor=cyan,
    citecolor=blue,
}

\AtBeginEnvironment{align}{\setcounter{subeqn}{0}}
\newcounter{subeqn} \renewcommand{\thesubeqn}{\theequation\alph{subeqn}}%
\newcommand{\subeqn}{%
  \refstepcounter{subeqn}
  \tag{\thesubeqn}
}

\begin{document}

\title[]{Simulation of convective transport during frequency chirping of a TAE using the MEGA code}

\author{H. Hezaveh$^1$, Y. Todo$^2$, Z. S. Qu$^1$, B. N. Breizman$^3 $ and M. J. Hole$^{1,4}$}

\address{$^1$ Mathematical Sciences Institute, The Australian National University}
\address{$^2$ National Institute for Fusion Science, National Institutes of Natural Sciences, Toki, Gifu 502-5292, Japan}
\address{$^3$ Institute for Fusion Studies, The University of Texas at Austin, Austin, TX 78712, United States of America}
\address{$^4$ Australian Nuclear Science and Technology Organisation, Locked Bag 2001, Kirrawee DC, NSW, 2232, Australia}

\ead{hooman.hezaveh@anu.edu.au}
\date{today}

\begin{abstract}
We present a procedure to examine energetic particle phase-space during long range frequency chirping phenomena in tokamak plasmas. To apply the proposed method, we have performed self-consistent simulations using the MEGA code and analyzed the simulation data. We demonstrate a travelling wave in phase-space and that there exist specific slices of phase-space on which the resonant particles lie throughout the wave evolution. For non-linear evolution of an $n=6$ toroidicity-induced \Alfven eigenmode (TAE), our results reveal the formation of coherent phase-space structures (holes/clumps) after coarse-graining of the distribution function. These structures cause a convective transport in phase-space which implies a radial drift of the resonant particles. We also demonstrate that the rate of frequency chirping increases with the TAE damping rate. Our observations of the TAE behaviour and the corresponding phase-space dynamics are consistent with the Berk-Breizman (BB) theory.
\end{abstract}

%
%
%
%
\ioptwocol

\section{Introduction}
\label{sec:intro}
The physics of energetic particles (EPs) plays an essential role in fusion plasmas. It has very attractive diagnostic applications but, on the other hand, it involves the possibility of unacceptably fast particle losses. A famous example is the destabilization of weakly damped  plasma waves inside the gaps of the shear \Alfven continuum, which entails redistribution or ejection of EPs either through diffusive transport or a convective transport where an isolated resonance moves radially like a bucket that carries resonant particles. The latter is associated with long range frequency chirping and has been observed for a variety of modes in experiments \cite{Gryaznevich2000,Maslovsky2003,Maslovsky2003b,Fredrickson2006,Matthew2019}. Refs. \cite{Duong1993,Munoz2011,Nabais2010,Heidbrink2014,Fredrickson2009,Podesta2009} show a correlation between wave-particle resonant interactions and fast ion loss and redistribution.

The formation of coherent structures in fast electrons phase-space was observed in non-linear simulations of a 1D electrostatic wave in Ref. \cite{Berk1997}. These structures (holes and clumps) are BGK-type modes with a chirping frequency. They evolve adiabatically and carry the trapped particles. Non-perturbative adiabatic models \cite{Boris2010,Nyqvist2012,Nyqvist2013,Hezaveh2017,Hezaveh2021} suggest the slow evolution of a Langmuir wave as a 1D paradigm of the more general wave-particle interactions in realistic geometries. In Refs. \cite{Hezaveh2020,Wang2018}, the theory has been extended to tokamak applications where the frequency chirping of \Alfvenic perturbations are studied. Ref. \cite{haowang2013} demonstrates the formation of holes and clumps during frequency chirping of the $n=0$ EGAM modes, where the toroidal momentum $(P_{\varphi})$ of the EPs is conserved in the presence of the electrostatic perturbations. The impact of EP beta value $(\beta_{\text{EP}})$ on chirping of a TAE mode was studied in Ref. \cite{xianquWang2020} and it has been shown that as the frequency of the wave changes, the dominant perturbation occurs at different slices of phase-space $(P_{\varphi}$ vs $E$ with $\mu=\text{const}$). In Ref. \cite{White2019}, the phase-space dynamics of EPs are studied during the long range frequency chirping of a TAE with a fixed eigenfunction, where phase-space slices are determined using two constants of motion, namely $\mu$ and $C=\omega_{\text{TAE}}P_{\varphi}-nE$ (see Refs. \cite{White_C,Briguglio2017}) with $\mu$, $\omega_{\text{TAE}}$, $n$ and $E$ being the magnetic moment, linear eigenfrequency, toroidal mode number and the EP energy, respectively. Still the question of how the chirping wave transports particles in phase-space deserves more detailed analysis. Technically speaking, best suited constants of motion for EPs dynamics need to be defined as the frequency evolves. 

In this work, we describe an appropriate procedure to observe the EPs dynamics on phase-space sub-slices using the adiabatic approximation for the frequency chirping of a TAE mode. Subsequently, we validate this method by applying the corresponding analysis to the results of EP simulations with the MEGA code \cite{Todo1998,todo2006}. We also show that the rate of frequency chirping is directly related to the damping rate of the modes in the bulk plasma. We demonstrate the latter by altering the dissipation coefficients when the mode has already evolved into chirping regime. In order to increase the resolution in phase-space, we have added test particles to the code. These particles respond to the perturbed field but do not contribute to the EP current self-consistently as the TAE evolves.

The rest of the paper is structured as follows: In section \ref{sec:MEGA}, we introduce a set of equations implemented in the hybrid MEGA code. Section \ref{sec:phase-space_anlys} describes the appropriate coordinates and constants of motion needed to analyse the guiding centre dynamics of EPs in phase-space during the non-linear frequency chirping. This involves canonical action-angle variables. Subsequently, we apply our phase-space analysis to the simulation data of the MEGA code in section \ref{sec:res} and report on the evolution of the TAE parameters. We identify resonant particles and exhibit their convective transport in phase-space. Section \ref{sec:summary} contains concluding remarks.  

\section{The simulation model in MEGA}
\label{sec:MEGA}
We simulate the evolution of the energetic particle driven mode within a hybrid model implemented in the MEGA code, where the bulk plasma particles are described as a fluid by the non-linear MHD equations and the fast particles are treated in a drift-kinetic approach. The MEGA code solves the following set of equations:
\\*
The momentum balance equation given by
\begin{eqnarray}
\rho\pdv{\bm{v}}{t} = - \rho\bm{v} \cdot \bm{\nabla} \bm{v} -\bm{\nabla} p + \left(\frac{1}{\mu_0} \bm{\nabla} \cross \bm{B} - \bm{j}_{\alpha} \right )   \nonumber \\
\cross \bm{B} +\frac{4}{3}\bm{\nabla}\left(\nu \rho \bm{\nabla} \cdot \bm{v}\right) - \bm{\nabla} \cross \left ( \nu \rho \bm{\nabla} \cross \bm{v}  \right),
\label{eq:mom}
\end{eqnarray}
where $\bm{j}_{\alpha}$ denotes the EP current, $\nu$ is the viscosity coefficient and $\rho$ and $p$ are the density and scalar pressure of the bulk plasma, respectively.
\\*
The continuity equation for the bulk plasma 
    \begin{equation}
        \pdv{\rho}{t} = -\bm{\nabla} \cdot \left(  \rho\bm{v}\right) + \nu_{n}\Delta \left (\rho - \rho_{0} \right ),
        \label{eq:continuity}
    \end{equation}
    where $\nu_n$ is the mass diffusivity. The energy balance equation for the evolution of the bulk plasma pressure
\begin{eqnarray}
\pdv{p}{t} &= - \bm{\nabla} \cdot \left ( p\bm{v} \right ) - \left ( \gamma-1 \right ) p \bm{\nabla} \cdot \bm{v} + \left ( \gamma-1 \right ) \\ \nonumber 
& \cross \left [\nu \rho \left ( \bm{\nabla} \cross \bm{v} \right )^2  + \frac{4}{3}\nu \rho \left ( \bm{\nabla} \cdot \bm{v} \right )^2  + \eta \bm{j} \cdot \left ( \bm{j} - \bm{j}_0 \right )  \right ] \\ \nonumber
& + \lambda \Delta \left ( p - p_0 \right ),
\label{eq:eng}
\end{eqnarray}
where $\gamma$ is the adiabatic constant and $\lambda$ represents the heat conductivity.
\\*
The set of Maxwell's equations and the Ohm's law given by
\begin{align}
&\pdv{\bm{B}}{t}  = -\bm{\nabla} \cross \bm{E}, \label{eq:max1} \refstepcounter{equation} \subeqn \\
&\bm{j} = \frac{1}{\mu_0} \bm{\nabla} \cross \bm{B}, \label{eq:ohm} \subeqn \\
&\bm{E} = -\bm{v} \cross \bm{B} +\eta \left (\bm{j} -\bm{j}_0  \right ),  \label{eq:max2} \subeqn
\end{align}
where $\eta$ represents resistivity. 

In the above equations, all the other quantities above are conventional. The subscript $0$ represents the equilibrium values of the parameters and the corresponding terms, as the source terms, have been used to enforce MHD equilibrium and compensate the diffusion and dissipation of the equilibrium fields. This set of equations is discretized using the method of finite difference and the fields are solved in an Eulerian scheme where the computational domain is gridded. 

The EPs are treated kinetically in a Lagrangian picture. A particle-in-cell method is applied to project the impact of EPs (EPs charge) on the grid points and update the fields in a self-consistent manner at each time step. The perturbation of the EPs, due to the wave, is  calculated using the $\delta f$ approach as the time evolution of the weight of each particle. This gives the following expression for the EPs current
\begin{eqnarray}
\bm{j}_{\alpha} =&\sum_{i=1}^{N}e Z_{\alpha} w_i  \left ( \bm{v}^{*}_{\parallel,i} +  \bm{v}_{B,i}   \right) S(x-x_i) \\ \nonumber
& - \bm{\nabla} \cross  \left [ \bm{b} \sum_{i=1}^{N} \mu_i w_i S \left ( x-x_i  \right ) \right ],
\label{eq:EP_current}
\end{eqnarray}
where the subscript $i$ represents the $i$th EP, $e Z_{alpha}$ is the charge of the EPs, $w_i$ is the weight, $\bm{v}_{B,i}$ is the drift due to the gradient of the magnetic field, $S$ is the shape factor and $\mu=E_{k} \left ( 1-\lambda^2  \right )/B$ is the magnetic moment with $E_{k}$, $\lambda$ and $B$ being the kinetic energy, pitch angle and the magnetic field at the guiding centre, respectively, $\bm{v}^{*}_{\parallel}$ contains the parallel velocity $v_{\parallel}$ to the magnetic field and magnetic curvature drift, and is given by
\begin{equation}
\bm{v}^{*}_{\parallel}= \frac{v_{\parallel}}{B^*}[\bm{B} + \rho_{\parallel} B \bm{\nabla} \cross \bm{b}],
\end{equation}
where $\rho_{\parallel}=\frac{mv_{\parallel}}{eZ_{\alpha}B}$ is the parallel gyro-radius and $B^*= B(1+\rho_{\parallel}\bm{b} \cdot \nabla \cross \bm{b})$ \cite{littlejohn1983}. It is noteworthy that $\bm{j}_{\alpha}$ does not contain $\bm{E} \cross \bm{B}$ drift due to quasi-neutrality \cite{Todo1998}. The EPs current is coupled to the MHD equations through Eq. \eqref{eq:mom}.

\section{Phase-space study}
\label{sec:phase-space_anlys}
In this section, we introduce canonical momenta that remain constant not only in the perturbative linear phase of the TAE evolution but also during the long range frequency chirping. Wee start from the the Littlejohn's Lagrangian \cite{Littlejohn} given by
\begin{equation}
L_{\text{littlejohn}}= e (\bm{A} + \rho_{\parallel} \bm{B})\cdot \bm{\dot{X}}+ \frac{m_i}{e}\mu\dot{\Omega}-H,
\label{eq:littlejohn_org}    
\end{equation}
where $e$ is the electron charge, $\bm{X}$ is the guiding centre position, $m_i$ is the ion mass, $\Omega$ is the gyro angle, $\bm{A}$ is the vector potential and $\bm{B}=\bm{\nabla} \cross \bm{A}$ and $H=\frac{1}{2}mv_{\parallel}^2+\mu B$ is the Hamiltonian.


For common choices of magnetic field line coordinates e.g. Boozer \cite{Boozer1981}, PEST \cite{Grimm1976}, Hamada \cite{Hamada1962} and etc, the guiding centre Lagrangian does not immediately reveal three canonical pairs of the Hamiltonian structure. This is due to the fact that the Lagrangian contains the time derivative of four variables as opposed to three. There have been several attempts to tackle this issue \cite{Boozer1981,White2014,Meiss1990} but each has its own disadvantages. In Ref. \cite{Meng}, the problem is resolved by introducing canonical angles, namely $(\theta_c,\xi_c)$, which give a new type of global coordinates called canonical straight field line coordinates. 


Using the new coordinates, a Legendre transformation can be implemented to find the equilibrium Hamiltonian 
\begin{equation}
H_0(P_{\theta_c},P_{\xi_c},P_{\Omega},\theta_c) = P_{\theta_c} \dot{\theta_c} + P_{\xi_c}\dot{\xi_c} + P_{\Omega}\dot{\Omega} - L_{\text{eq}},
\label{eq:equilibrium_hamiltonian_meng}    
\end{equation}
which describes the unperturbed guiding centre dynamics of EPs with
\begin{align}
&P_{\theta_c}  = e\psi + mv_{\parallel} b_{\theta_c}, \label{eq:P_theta_c} \refstepcounter{equation} \subeqn \\
&P_{\xi_c} = -e\chi + mv_{\parallel} b_{\varphi_c}, \label{eq:P_chi_c} \subeqn \\
&P_{\Omega} = \frac{m}{e}\mu.  \label{eq:P_omega} \subeqn
\end{align}
The set $\left ( P_{\theta_c},P_{\xi_c},P_{\Omega}\right )$ are the canonical momenta conjugated to $(\theta_c,\xi_c,\Omega)$. For this completely integrable system, the $\theta_c-$dependence of the Hamiltonian can be eliminated by using a canonical transformation to action-angle variables. In these variables, we have
\begin{equation}
H_0 = H_0 (P_{\tilde{\theta}_c},P_{\tilde{\xi}_c},P_{\tilde{\Omega}}),
\label{eq:Hamilton_AA}    
\end{equation}
where the action variables $(P_{\tilde{\theta}_c},P_{\tilde{\xi}_c},P_{\tilde{\Omega}})$ correspond to the angles $(\tilde{\theta}_c,\tilde{\xi}_c,\tilde{\Omega})$ that are linear functions of time in the unperturbed motion, i.e.,
\begin{align}
\dot{\tilde{\theta}}_c = \pdv{H_0}{P_{\tilde{\theta}_c}} =\omega_{\tilde{\theta}_c}, \refstepcounter{equation}	\label{eq:freq_1}	\subeqn \\
\dot{\tilde{\xi}}_c = \pdv{H_0}{P_{\tilde{\xi}_c}} =\omega_{\tilde{\xi}_c}.  \label{eq:freq_2}   \subeqn
\end{align}
To describe the perturbed motion of the particles, we write their total Hamiltonian $H_{\text{total}}$ as a sum of the unperturbed Hamiltonian $H_0$ and a perturbation $U$ associated with the wave. This gives
\begin{equation}
H_{\text{total}} = H_0 + U.
\end{equation}
We use the following representation for the perturbation $U$
\begin{equation}
U = \Sigma_{h,m} \phi_{m;n;h} \left(r;t\right)\e^{ih(m\theta_c+n\xi_c-\alpha\left(t\right))}.
\label{eq:uu}
\end{equation}
This representation corresponds to a single chirping wave formed and evolved as a BGK-type wave through excitation of sideband/secondary oscillations of a single eigenmode in an isolated resonance. We rewrite $U$ in terms of the action-angle variables of the unperturbed motion to have 
\begin{equation}
H_{\text{total}} = H_0 + U(P_{\tilde{\theta_c}}; P_{\tilde{\xi_c}}; P_{\tilde{\Omega}};p\tilde{\theta_c}+ h \left [ n\tilde{\xi_c}-\alpha\left(t \right) \right ] ).
\end{equation}
Here, $U$ is associated with an individual particle resonance, denoted by $l=\frac{p}{h}$, which includes several terms from expression \eqref{eq:uu} i.e. $U$ is a periodic function but not necessarily sinusoidal. The coefficients of the aforementioned expansion are the orbit-averaged mode amplitudes which represent the coupling strength (see Refs. \cite{Berk1995,Hezaveh2020}). For the dynamics governed by the total Hamiltonian given above, $P_{\tilde{\Omega}}$ is already a conserved quantity and since the Hamiltonian depends on a combination of $\tilde{\theta_c}$ and $\tilde{\xi_c}$, we have another immediate conservation law which makes the problem essentially one dimensional. This 1D description of wave-particle interaction can be represented by transferring the coordinates canonically to a frame co-moving with the chirping wave. A type-2 generating function for such a transformation is
\begin{equation}
G_2 \left ( \bm{q},\bm{p_{\text{new}}} , t  \right) = P_{1} \left [ l \tilde{\theta_c} + n \tilde{\xi_c} - \alpha \left ( t \right ) \right ] + P_{2} \tilde{\xi_c} + P_{3} \tilde{\Omega}.
\label{eq:gf2}
\end{equation}
It can be used to write the explicit expressions for the new variables and constants of motion as
\begin{equation}
\begin{split}
P_1 &= \frac{1}{l} P_{\tilde{\theta_c}}\\ P_2 &= P_{\tilde{\xi_c}} + \frac{n}{l} P_{\tilde{\theta_c}}\\  P_3 &= P_{\tilde{\Omega}}
  \end{split}
\ \ \ \ \ \
\begin{split}
Q_1 &= \zeta_{l} = l \tilde{\theta_c} + n \tilde{\xi_c} - \alpha \left ( t \right )\\ Q_2 &= \tilde{\xi_c}\\ Q_3 &= \tilde{\Omega}, 
 \end{split}
\label{eq:CT1}
\end{equation}
after which the new Hamiltonian takes the form
\begin{equation}
H_{\text{new}} = H_{\text{eq}}(P_1,P_2,P_3) + U(\zeta,P_1,P_2,P_3),
\label{eq:H_new}
\end{equation}
where $P_2$ and $P_3$ are constants of motion and a generalized momentum $(P_1)$ and its corresponding coordinate $(\zeta)$, to which the momentum is conjugated, constitute the dynamical variables. We thereby follow the EPs dynamics in $P_1-\zeta$ on sub-slices of $P_2=\text{const}$ and $P_3=\text{const}$. The distinctive feature of the chosen variables is that $P_2$ remains conserved as the frequency chirps.   

\begin{figure*}[t!]
\centering
\subfloat[]{
\includegraphics[scale=0.5]{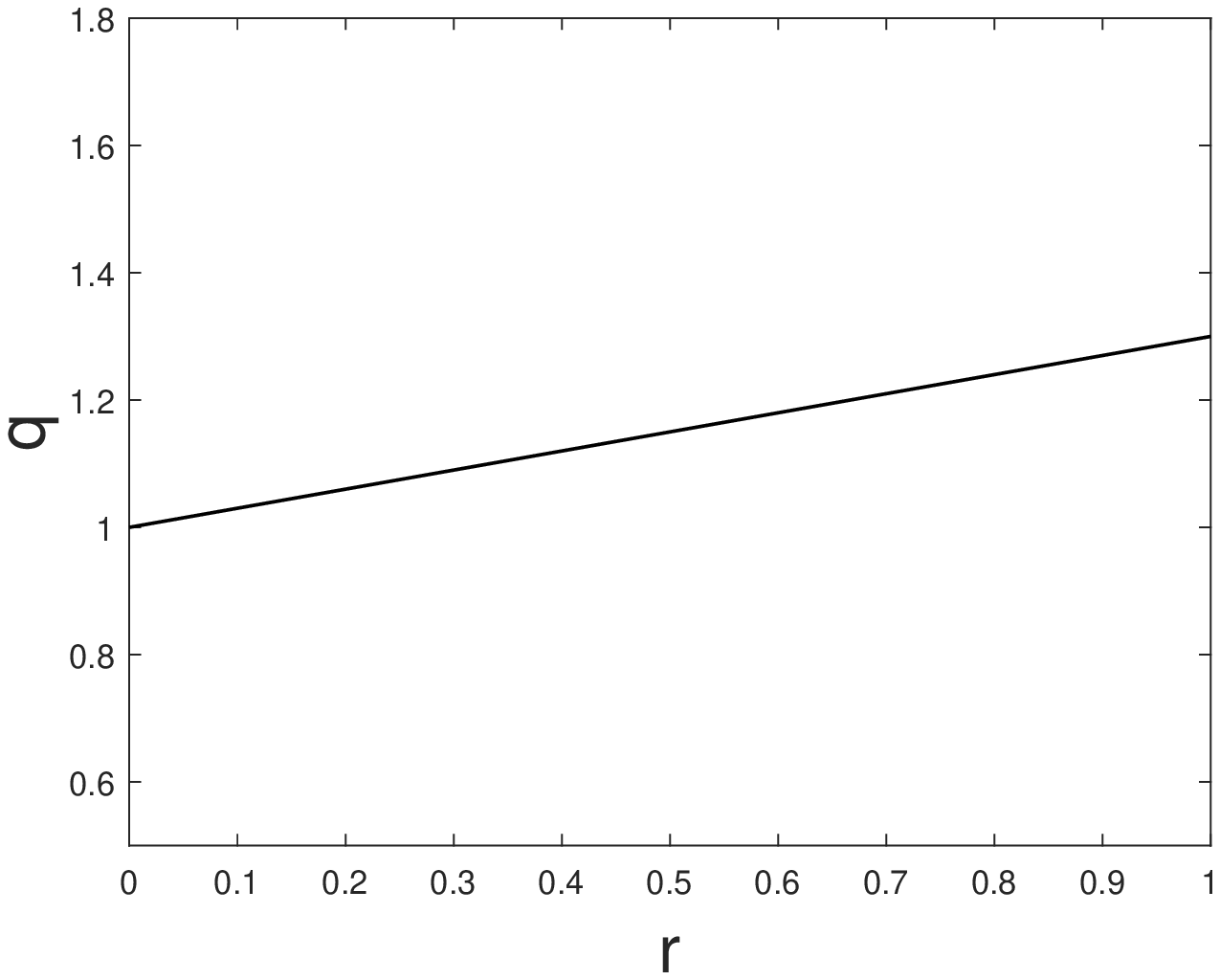}
\label{fig:q}
}
\subfloat[]{
\includegraphics[scale=0.5]{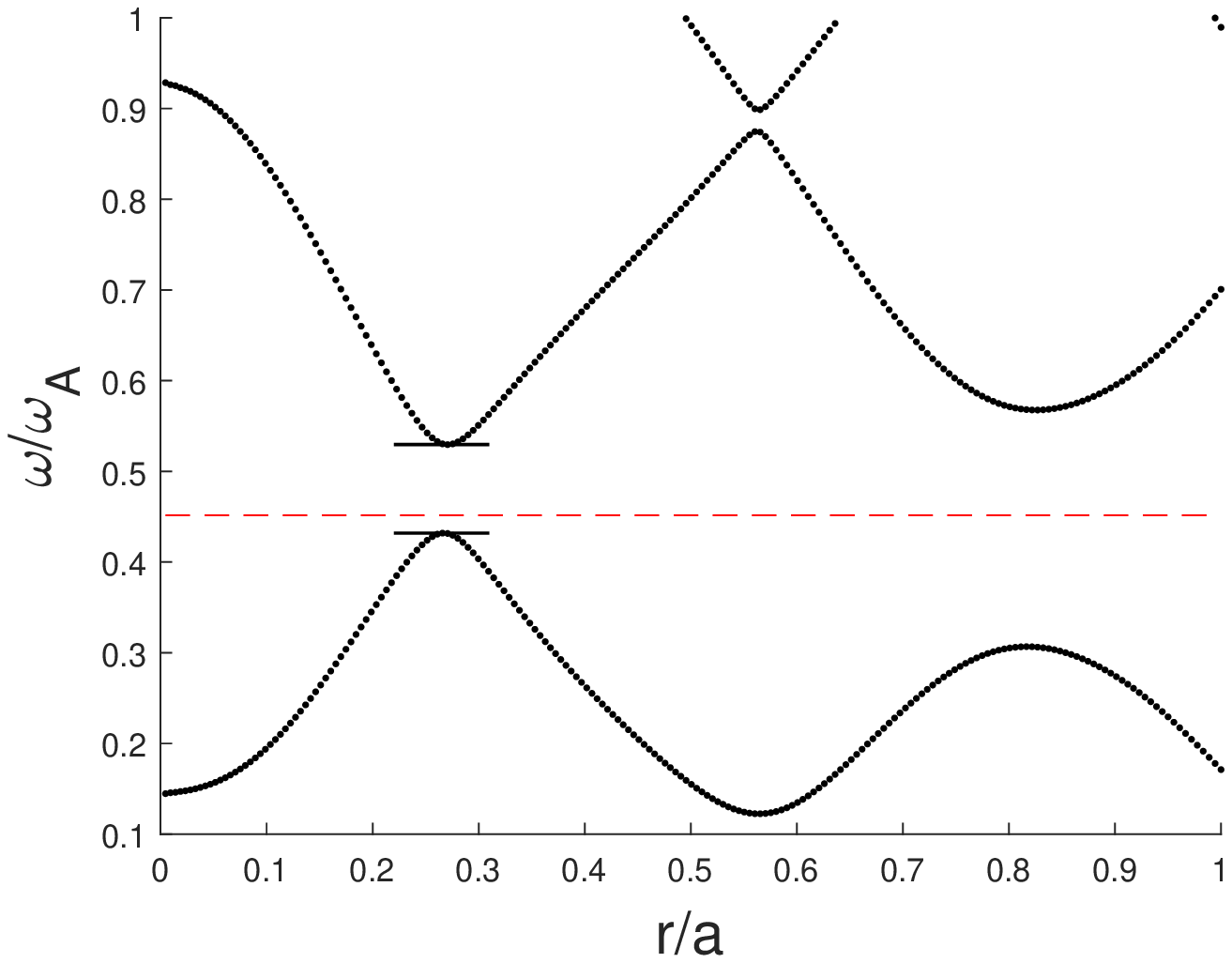}
\label{fig:cont}
}
\caption{(a) The safety factor and (b) the corresponding shear \Alfven continuum in a circular cross section configuration for $n=6$. The red dashed line represents the linear frequency of the toroidal \Alfven eigenmode.}
\label{fig:evl_SR}
\end{figure*}

\begin{figure*}[b!]
\centering
\subfloat[]{
\includegraphics[scale=0.52]{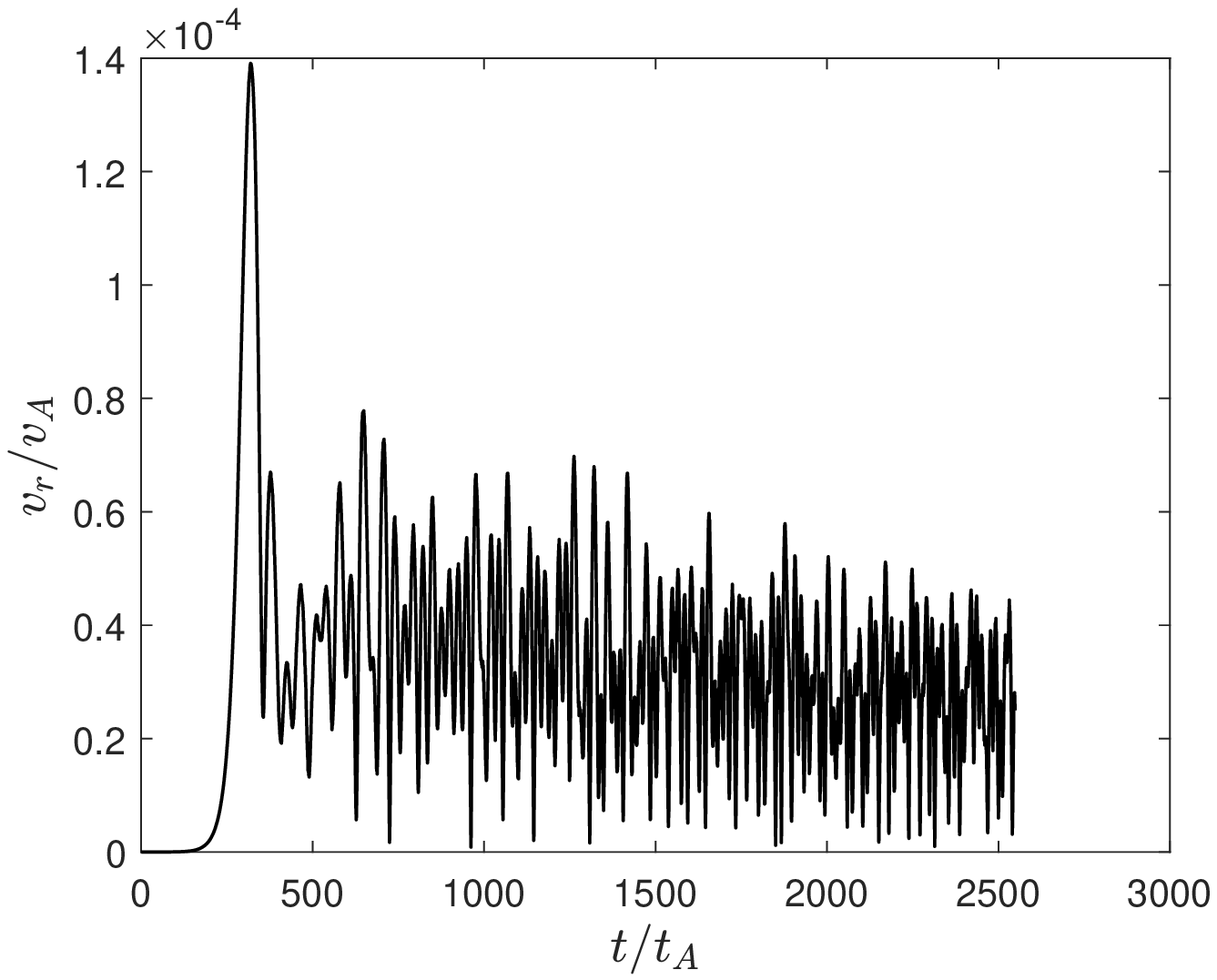}
\label{fig:V_r_norise}
}
\subfloat[]{
\includegraphics[scale=0.52]{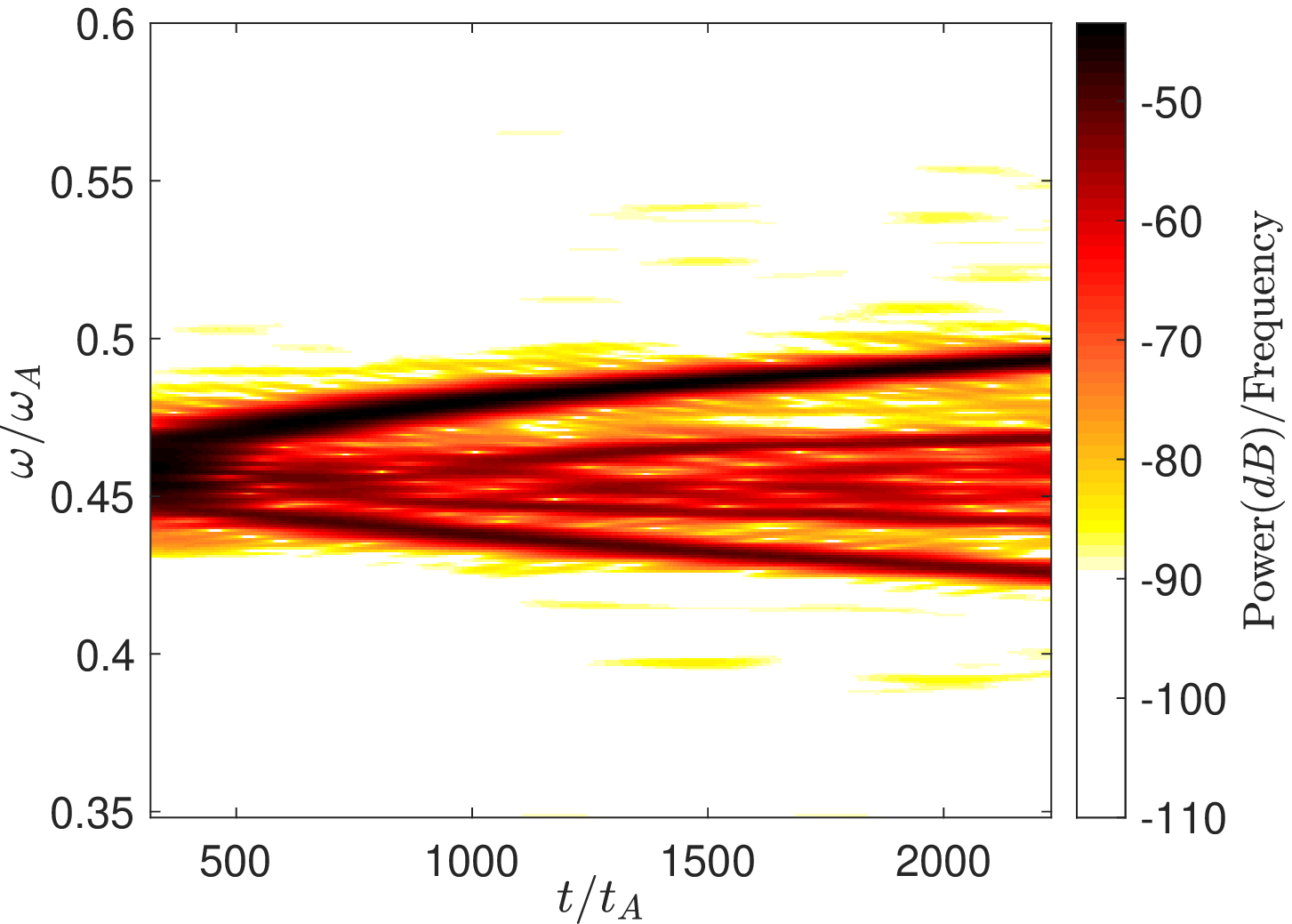}
\label{fig:spec_norise}
}
\caption{Time evolution of the TAE envelope (a) and the frequency spectrogram (b) for $n=6$, $m=6$ oscillations at $r/a=0.27$. The y-axis of panel a shows the normalised radial component of the plasma velocity. The color bar of panel b represents an estimate of the short-term time-localized frequency content of $\cos$ component of $v_r$. The dissipative coefficients are kept the same as given in expression \eqref{eq:dis_coef} throughout the mode evolution.}
\label{fig:evolve_norise}
\end{figure*}

So far, we have introduced proper coordinates for our phase-space analysis, and the next step is to identify how the six-dimensional coordinate transformation of $(P_{\theta_c},P_{\xi_c},P_{\Omega},\theta_c,\xi_c,\Omega) \to (P_{\tilde{\theta_c}},P_{\tilde{\xi}_c},P_{\tilde{\Omega}},\tilde{\theta}_c,\tilde{\xi}_c,\tilde{\Omega})$ is carried out. We do that by relating the EPs frequencies to $P_{\tilde{\theta_c}}$ and $P_{\tilde{\xi}_c}$ using Eqs. \eqref{eq:freq_1}, \eqref{eq:freq_2} and \eqref{eq:Hamilton_AA}.


Given \eqref{eq:Hamilton_AA}, $H_0$ is known as the particle energy $(E)$, and $P_{\tilde{\xi}_c}$ and $P_{\tilde{\Omega}}$ are also known quantities and can be evaluated using \eqref{eq:P_chi_c} and \eqref{eq:P_omega}, respectively, for $\xi_c$ and $\Omega$ being ignorable coordinates in $H_0$ of \eqref{eq:equilibrium_hamiltonian_meng}. Hence, Eq. \eqref{eq:Hamilton_AA} can be inverted to write
\begin{equation}
P_{\tilde{\theta}_c}=P_{\tilde{\theta}_c}(H_0=E,P_{\tilde{\xi}_c},P_{\tilde{\Omega}}).
\label{eq:p_theta}
\end{equation}
We use the following procedure to implement this inversion. For a slice of $\mu=\text{const}$, we write
\begin{equation}
P_{\tilde{\theta}_c}=G(E,P_{\tilde{\xi}_c}),
\label{eq:p_theta_fit}
\end{equation}
where $G$ is a 2D polynomial of $\sqrt{E}$ and $P_{\tilde{\xi}_c}$. The reason we take $G$ as a function of $\sqrt{E}$ is that in this work we focus on the highly passing particles $(\mu=0)$, as in a neutral beam injection (NBI) scenario, for which $\omega_{\tilde{\theta}_c}=\frac{v_{\parallel}}{qR_0} \propto \sqrt{E}$ and $\omega_{\tilde{\xi}_c}=\frac{v_{\parallel}}{R_0}\propto \sqrt{E}$. Therefore, we have $H_0(G(E,P_{\tilde{\xi}_c}),P_{\tilde{\xi}_c})$. Applying the derivative operator to both sides of \eqref{eq:p_theta_fit} with respect to $P_{\tilde{\theta}_c}$ and $P_{\tilde{\xi}_c}$ gives  
\begin{align}
&\pdv{G}{E} = \frac{1}{ \hat{\omega}_{\tilde{\theta}_c} } \text{and} \refstepcounter{equation}	\label{eq:}	\subeqn \\
&\pdv{G}{P_{\tilde{\xi}_c}} = -\pdv{G}{E}\hat{\omega}_{\tilde{\xi}_c},  \label{eq:}   \subeqn
\end{align}
respectively, where Eqs. \eqref{eq:freq_1} and \eqref{eq:freq_2} are used and $\hat{}$ denotes the frequencies calculated using the fitting function $G$. To fit $G$, we use the method of least squares with the following minimization function 
\begin{equation}
M = \frac{1}{N} [\sum_{i=1}^{N} (\frac{1}{\hat{\omega}_{\tilde{\theta}_c}}-\frac{1}{\omega_{\tilde{\theta}_c}})^2 +    \sum_{i=1}^{N} (\frac{1}{\hat{\omega}_{\tilde{\xi}_c}}-\frac{1}{\omega_{\tilde{\xi}_c}})^2 ],
\label{eq:min_func}
\end{equation}
where $N$ is the total number of EPs on a  $\mu=\text{const}$ slice. In order to evaluate $M$, the equilibrium frequencies $(\omega_{\tilde{\theta}_c},\omega_{\tilde{\xi}_c})$ must be determined from simulation. These are computed by tracing particle trajectories for different $P_{\tilde{\xi}_c}$ and $E$. Once known, the polynomial coefficients of $G$ are varied until Eq. \eqref{eq:min_func} is minimised. This determines $G$. 

\begin{figure*}[t!]
\centering
\subfloat[]{
\includegraphics[scale=0.52]{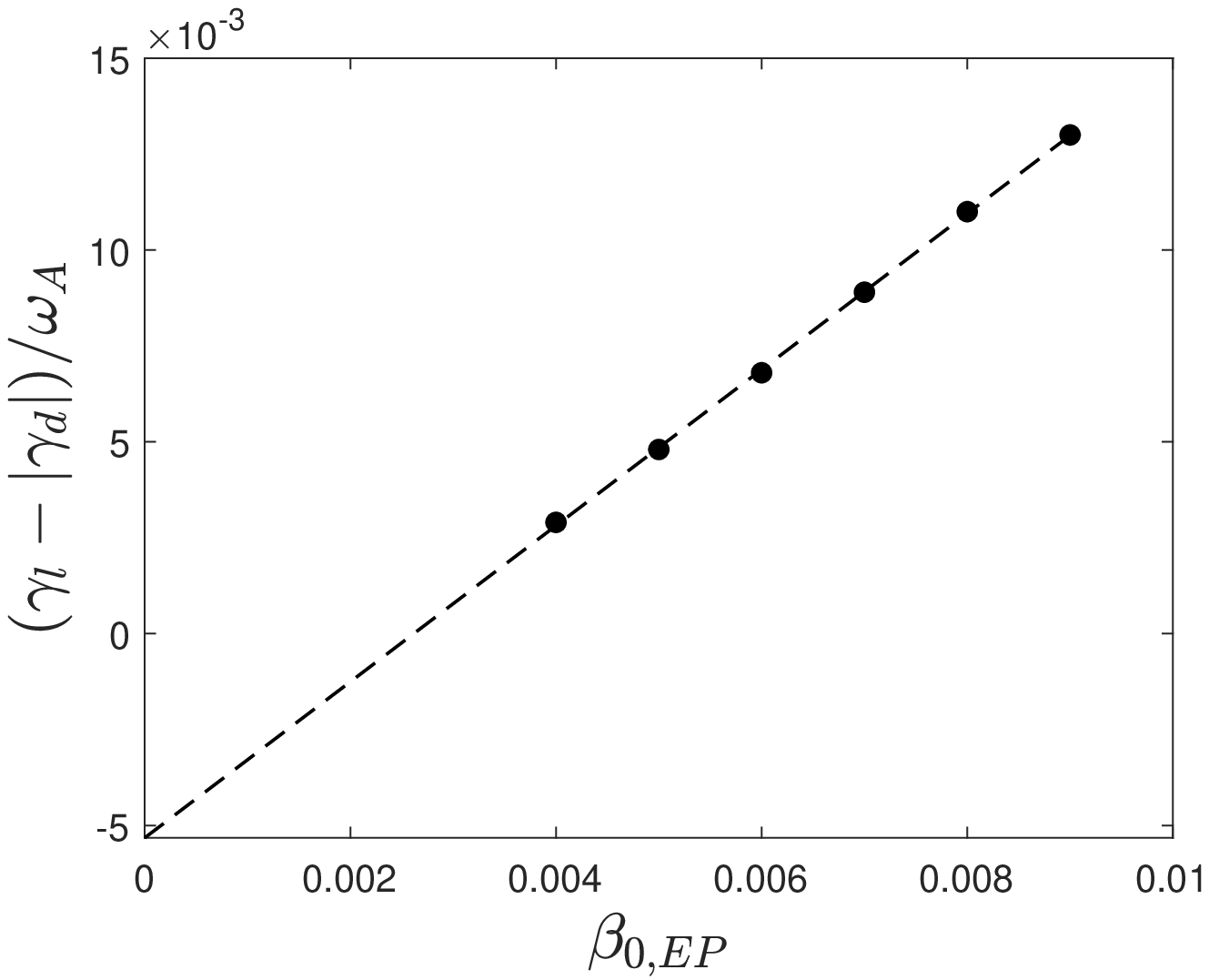}
\label{fig:gamma_d}
}
\subfloat[]{
\includegraphics[scale=0.52]{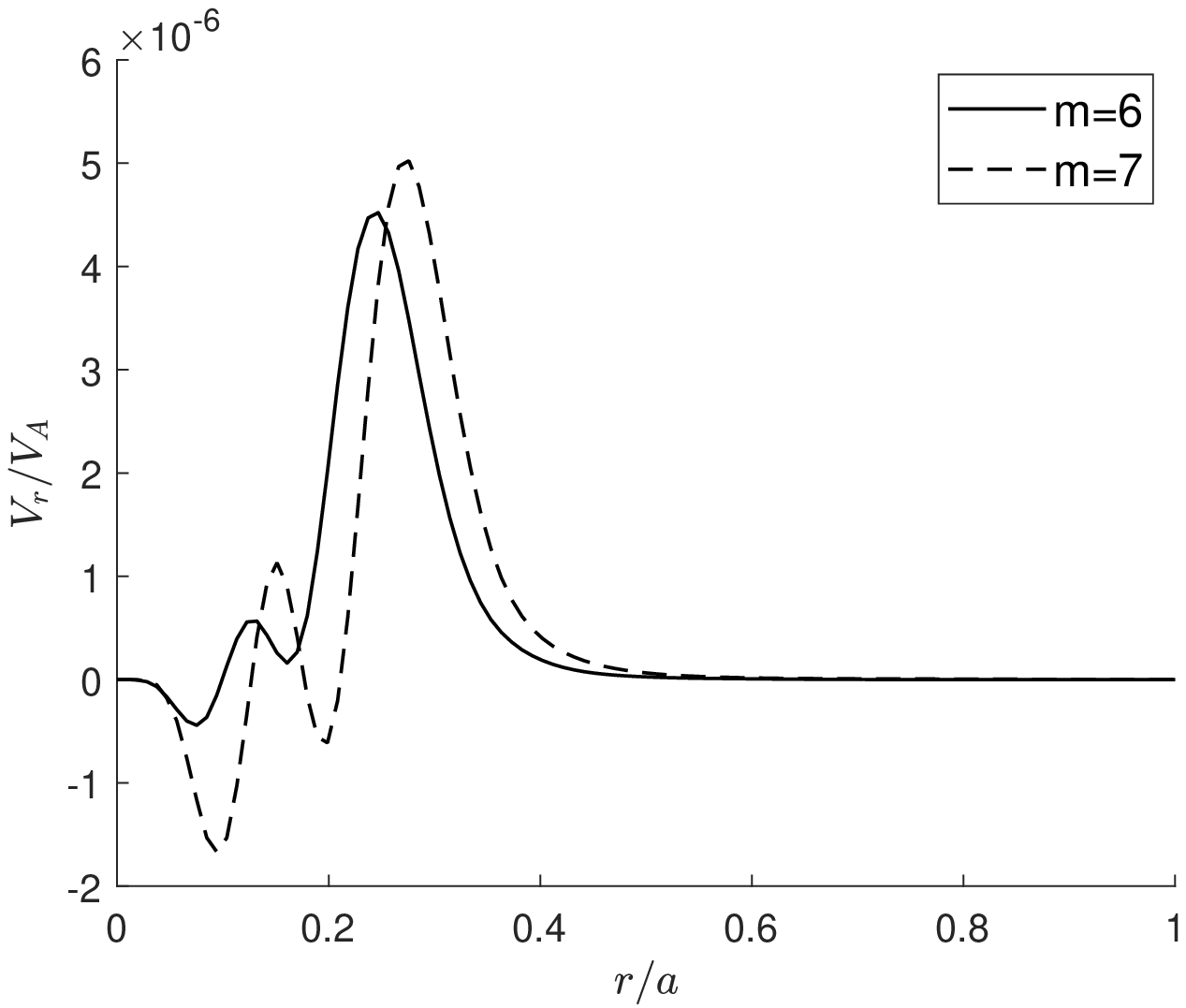}
\label{fig:radial_profile}
}
\caption{(a) A scan of the net growth rate versus EP pressure on axis from simulation data (the black circles) and a linear fit to the data (dashed line) and (b) the structure of the radial component of the bulk plasma velocity $(v_r)$, normalised to the \Alfven velocity on the axis $(v_A)$, versus the normalised minor radius at $t/t_A=217.6$. Here, $t_A$ is the \Alfven time on the axis}
\label{fig:evolve}
\end{figure*}

Considering the set $(R,z,\varphi)$ as the cylinderical coordinate, we consider $\tilde{\theta}_c=0$ and $\tilde{\xi}_c=\varphi$ on the $z=0$ plane with largest $R$ where we also record the particle data. On this plane, the canonical angles $(\theta_c,\xi_c)$ equal geometrical angles (see \cite{Meng} and Eq.23 of Ref.\cite{Zhisong2014}). As a convenient choice, this plane can also be used to show $P_{\tilde{\xi}_c}=P_{\varphi}$, where $P_{\varphi}$ is the toroidal angular momenta conjugated to $\varphi$.       
\\

The above approach gives an essentially 1D representation of the wave-particle interaction using phase-space plots in $P_1$-$\zeta$ space. A notable advantage of this method is that $P_2$ is conserved even when the frequency experiences long deviation from the initial eigenfrequency. This has important implications when resolving the question of whether the EPs trapped inside the chirping wave are carried with the wave (consistent with the adiabatic theory of frequency chirping) or different particles are perturbed by the wave as the frequency chirps.


\begin{figure*}[b!]
\centering
\subfloat[]{
\includegraphics[scale=0.52]{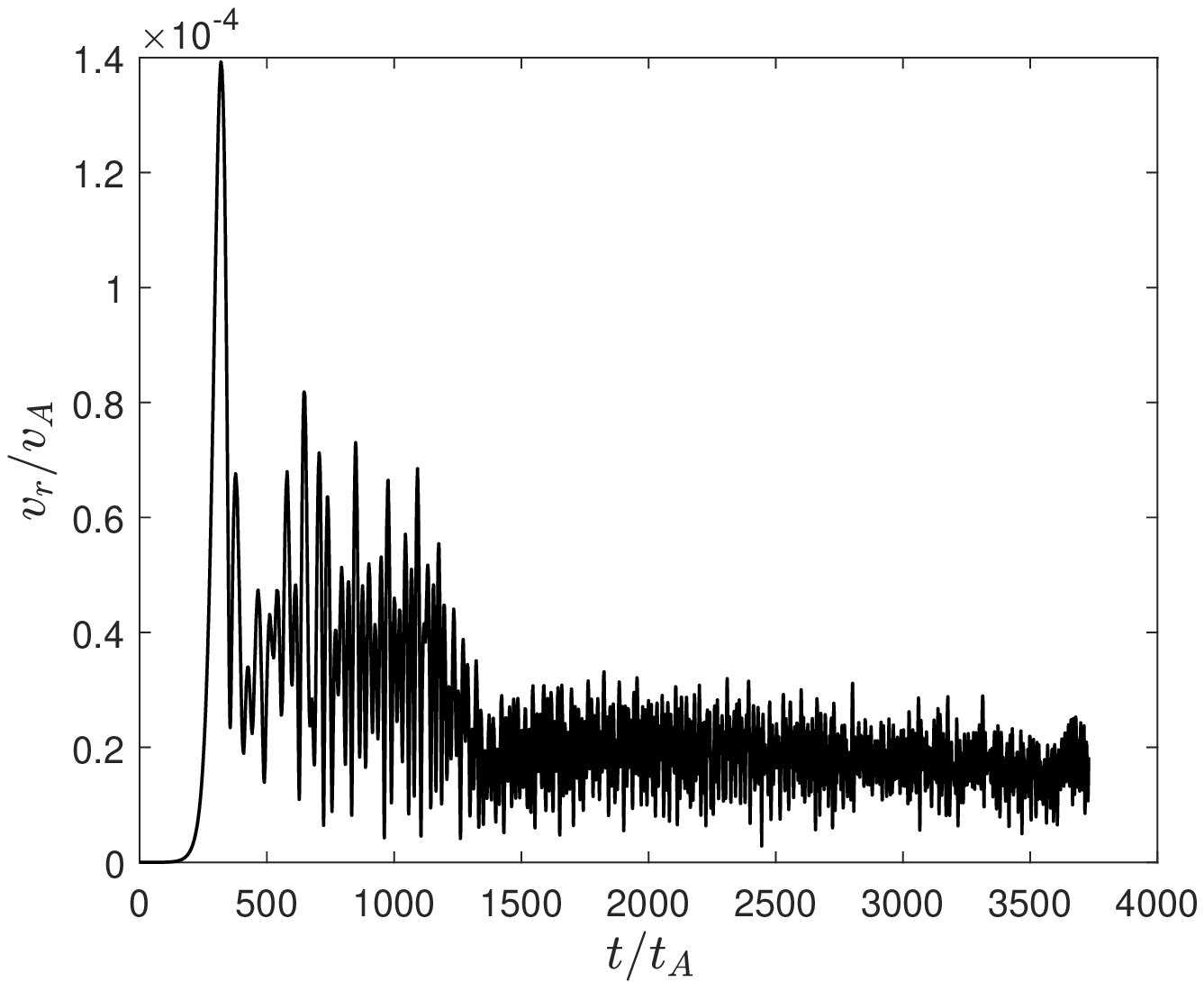}
\label{fig:V_r}
}
\subfloat[]{
\includegraphics[scale=0.52]{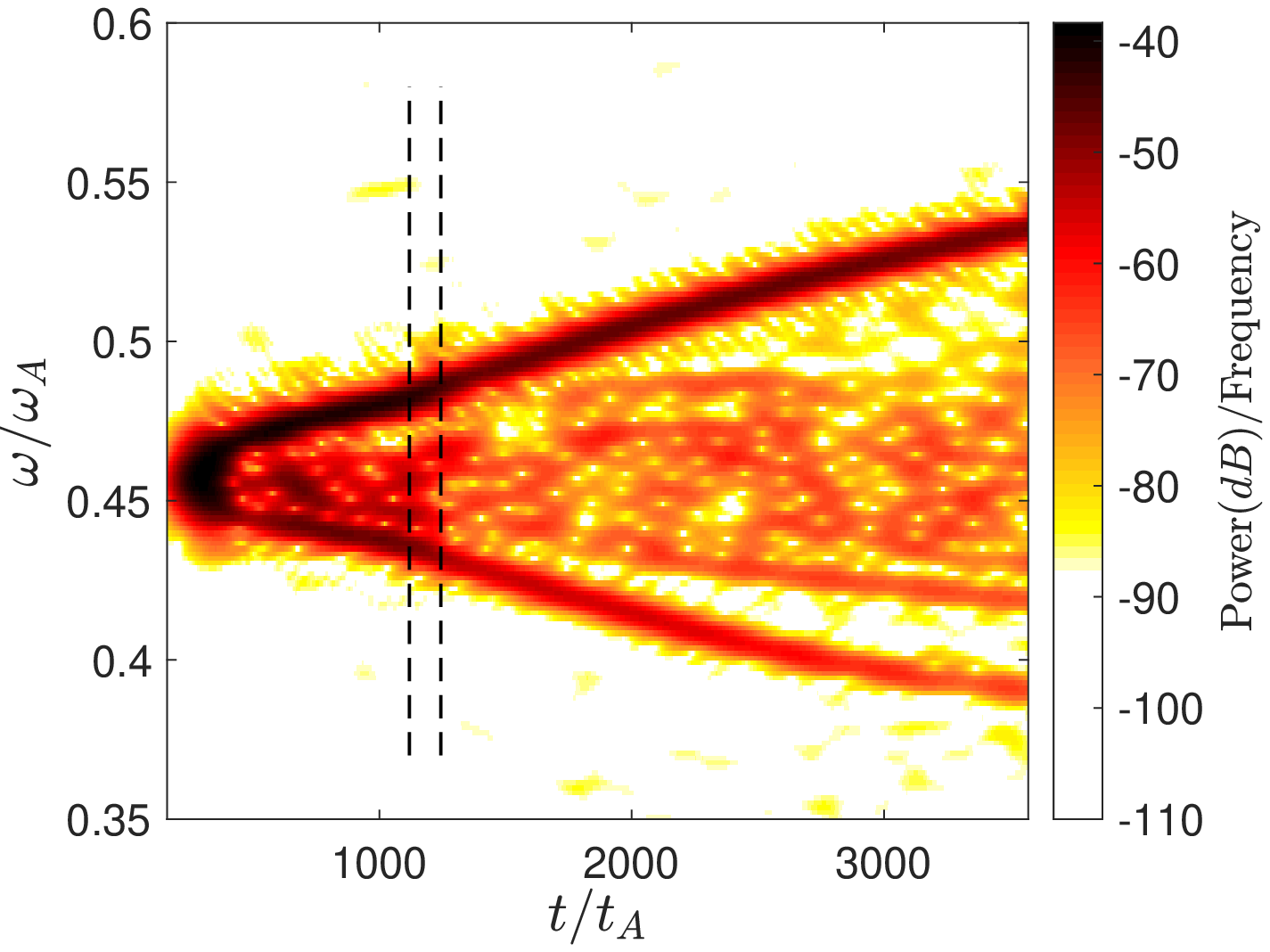}
\label{fig:spec_rise}
}
\caption{Time evolution of the TAE envelope (a) and the frequency spectrogram (b) for $n=6$, $m=6$ oscillations at $r/a=0.27$. The two vertical dashed lines on panel b denote the times, namely $t/t_A=1119.1$ and $1243.4$, at which the damping coefficients has been increased. The color bar represents an estimate of the short-term time-localized frequency content of $\cos$ component of $v_r$.}
\label{fig:evolve}
\end{figure*}

\section{Analysis of the simulations}
\label{sec:res}

The MEGA code uses an equilibrium configuration  constructed  by a Grad-Shafranov solver for a given q-profile. In this work, we use a linear q-profile depicted in \fref{fig:q}. We choose the inverse aspect ratio $\epsilon=3.2$. The density and pressure are uniform throughout the plasma. The corresponding shear \Alfven continuum for $n=6$ is plotted in \fref{fig:cont}. The accumulation points of the first gap are located at $r/a=0.26$. For a TAE, the q-profile has a rational value at the cylindrical cross-over points, where $q=(2m+1)/2n$. As shown in \fref{fig:q}, the first gap corresponds to $m=6$ coupled to $m=7$, and the second gap located at $r/a=0.82$ corresponds to $m=7$ that is coupled to $m=8$. The equilibrium phase-space density of EPs is initialized using a slowing down distribution given by
\begin{equation}
F_{\text{eq},\alpha} = \frac{\kappa}{E^3 + E_{\text{crt}}^3} [1 + \erf(\frac{E_0-E}{\Delta E})] \exp(-\frac{\langle\psi \rangle}{\Delta\psi}),
\label{eq:}
\end{equation}
where $E_{\text{crt}}$ and $E_0$ represent the critical and birth energies of the alpha particles, respectively, $\psi$ is the poloidal magnetic flux, $\langle\rangle$ denotes an averaged quantity, $\Delta E$ and $\Delta \psi$ specify the characteristic width of the equilibrium phase-space density in energy and $\psi$, respectively. For the purpose of this work, the values are set as $E_0=1.44E_A$, $E_{\text{crt}}=0.25E_A$, $\Delta E = 0.0144 E_A$ and $\Delta \psi=0.148\psi_{\text{max}}$, where $\psi_{\text{max}}$ is the maximum value of $\psi$ and $E_A=\frac{1}{2}mv_A^2 $ with $v_A$ being the \Alfven velocity at the centre of the plasma. The EPs pressure is set to give an EP beta value of $\beta_{0,EP}=0.6\%$ on the magnetic axis. The damping coefficients are
\begin{equation}
\nu=\eta=0.3 \cross 10^{-7} v_{A} R_{0}, \hspace{1cm} \nu_n=\lambda=0.  
\label{eq:dis_coef}
\end{equation}

\begin{figure*}[b!]
\centering 
\subfloat[]{\includegraphics[height=0.36\textwidth]{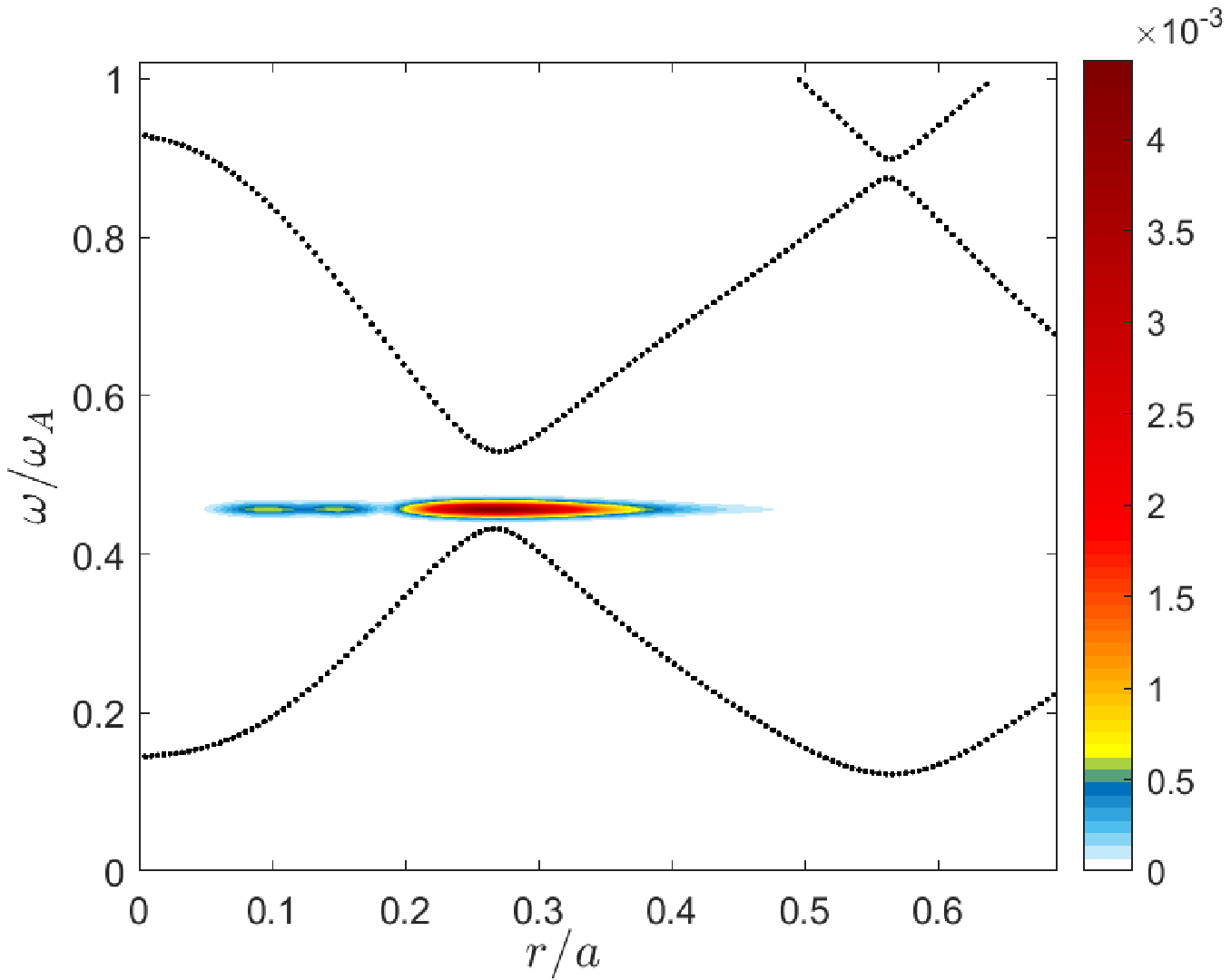} \label{fig:spec_rd1}} \qquad
\subfloat[]{\includegraphics[height=0.36\textwidth]{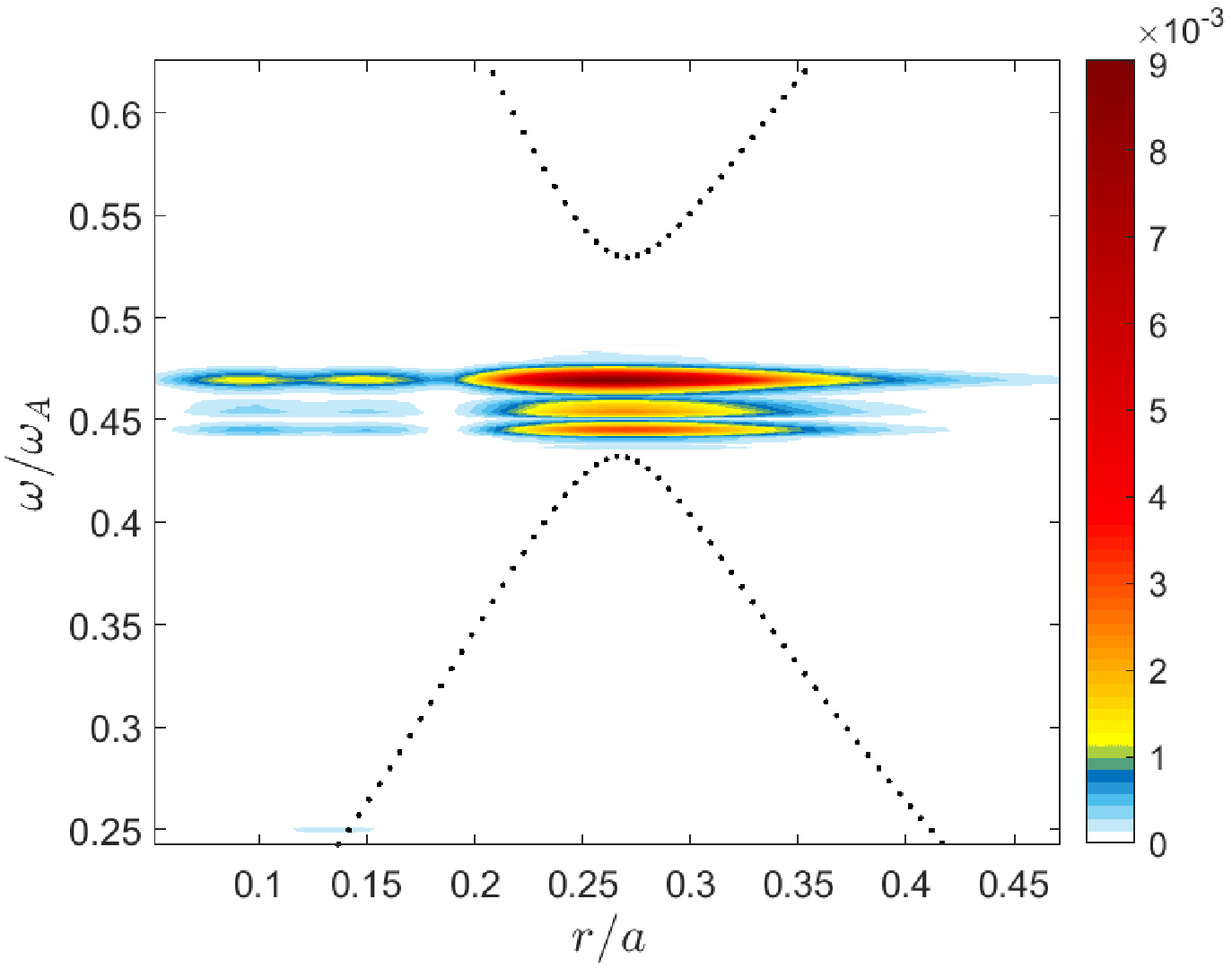}\label{fig:spec_rd2}}  
\\
\subfloat[]{\includegraphics[height=0.36\textwidth]{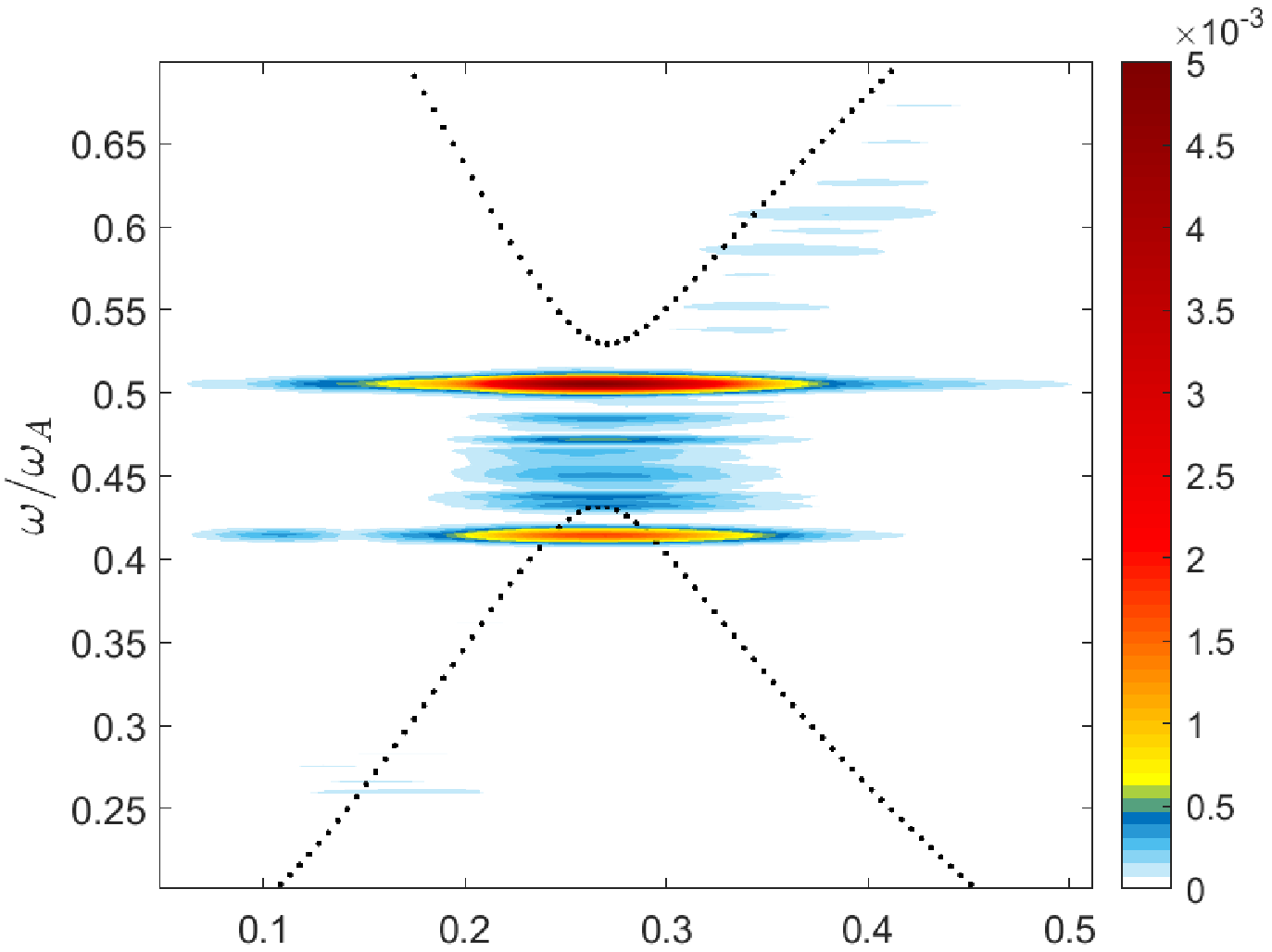}\label{fig:spec_rd3}}  \qquad
\subfloat[]{\includegraphics[height=0.36\textwidth]{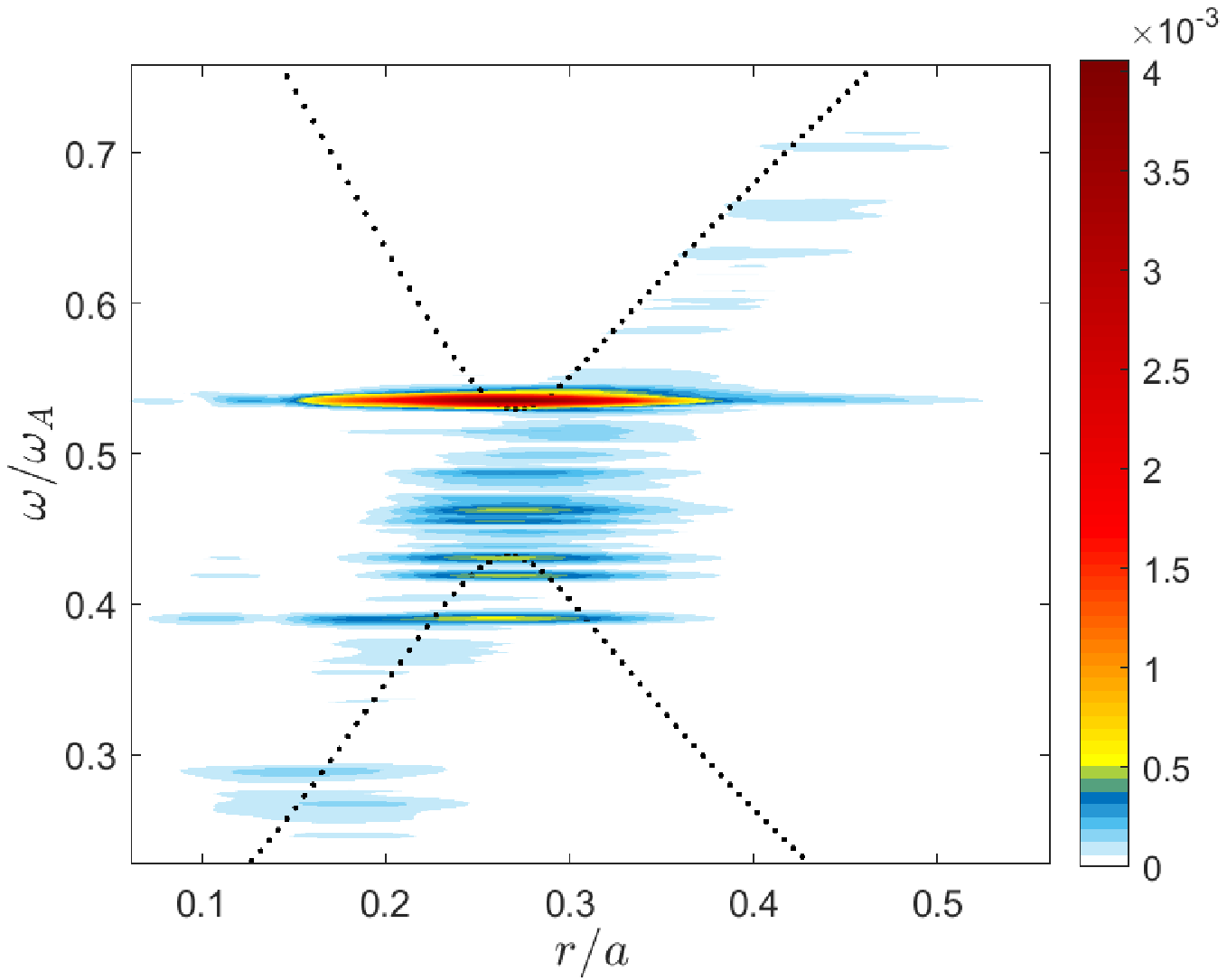}\label{fig:spec_rd4}}  
\caption{Frequency spectrum across the radial coordinate at different time slices. The color bar represents the absolute value of power(dB) per frequency. Panels (a), (b), (c) and (d) correspond to $t/t_A=196.1$, $t/t_A=469.7$, $t/t_A=1713.1$ and $t/t_A=3570.4$ respectively. The dotted line shows the Shear \Alfven continuum.}
\label{fig:mode_structure}
\end{figure*}

\subsection{Evolution of the driven eigenmode}

By solving the initial value problem with the MEGA code we find that the dominant perturbation is a TAE excited above the lower tip of the first gap with a linear frequency $\omega_{\text{TAE}}/\omega_A=0.4553$, where $\omega_A$ is the \Alfven frequency on the axis. \Fref{fig:evolve_norise} shows the evolution of the amplitude and the frequency of the mode. The absolute value of the plasma radial velocity $v_r$ is depicted in \fref{fig:V_r_norise} as a function of time. Using an exponential fit, we find that the net growth rate of the mode is $(\gamma_l - \abs{\gamma_d})/\omega_A=0.0067$. Similarly, we perform a scan of the net growth rate over $\beta_{0,EP}$ to find the damping rate $(\gamma_d)$ of the mode. 
This is depicted in \fref{fig:gamma_d} where a linear polynomial, fitted to the simulation data, identifies the intercept with the vertical axis. This gives a damping rate of $\gamma_{d}/\omega_A = 0.0053$. Subsequently, the linear growth rate of the TAE is $\gamma_l/\omega_A = 0.012$. Hence, in this simulation we have 
\begin{equation*}
\gamma_d/\gamma_l=0.44, \hspace{1cm} \gamma_l/\omega_{\text{TAE}}=2.64 \%. 
\end{equation*}
The two dominant radial profiles of the TAE corresponding to the poloidal mode numbers $m=6$ and $7$ are shown in \fref{fig:radial_profile}. We observe that the peak lies around the location of the first gap of the shear \Alfven continuum. 

\begin{figure*}[b!]
\centering
\subfloat[]{
\includegraphics[scale=0.57]{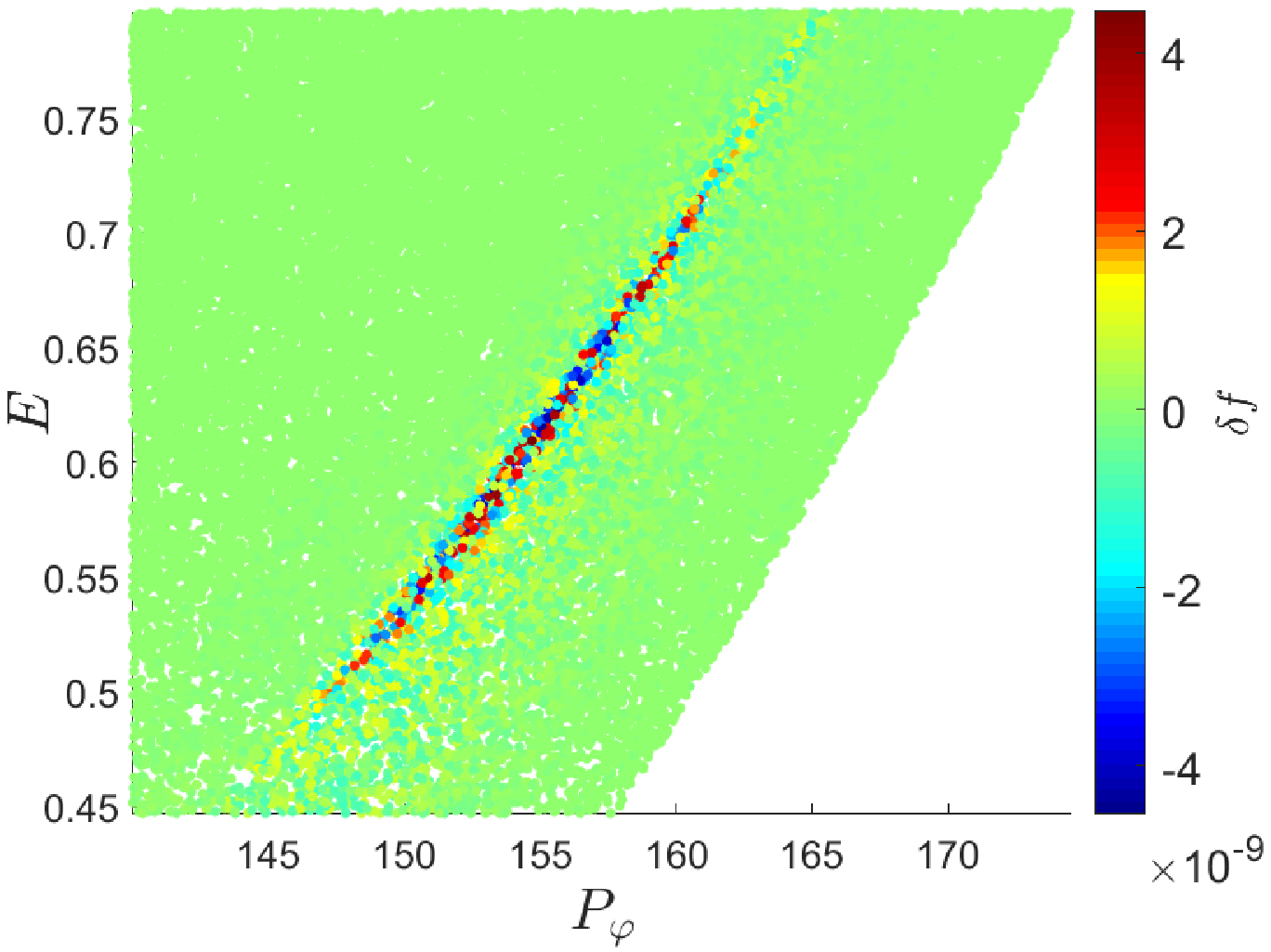}
\label{fig:phase-space1}
}
\subfloat[]{
\includegraphics[scale=0.57]{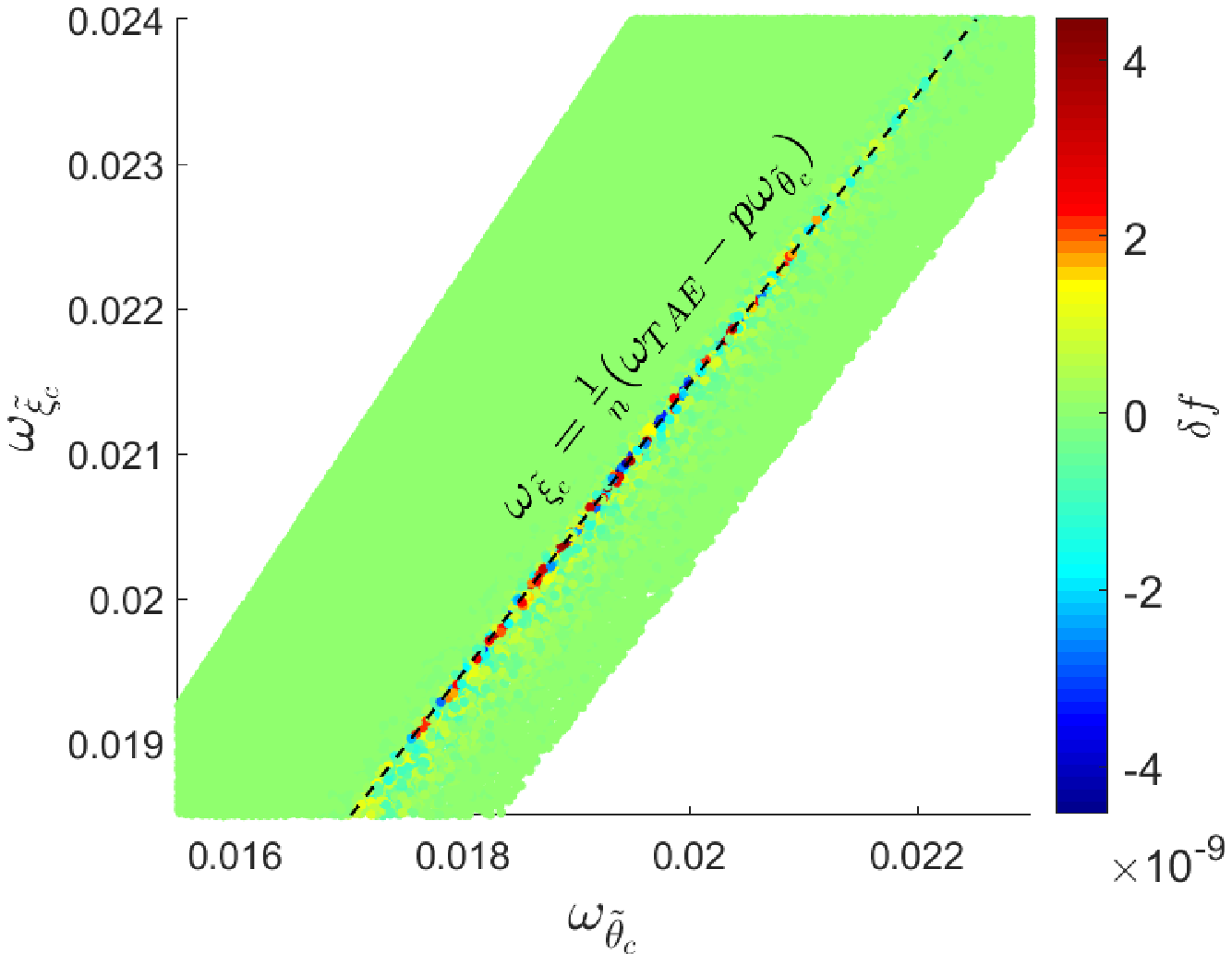}
\label{fig:phase-space}
}
\caption{The resonance curve (a) and the resonance line (b) in $E$ vs. $P_{\varphi}$ and $\omega_{\tilde{\xi}_c}$ vs. $\omega_{\tilde{\theta}_c}$ plane, respectively. Each panel shows a $\mu = 0$ slice of the phase-space for co-passing EPs. The color bar represents the particle weights (perturbed distributions). The dashed line is a fit using the resonance condition of \eqref{eq:res}.}
\label{fig:ps}
\end{figure*}

\Fref{fig:spec_norise} shows an evolving spectrum of the cosine part of $v_r$. It reveals the primary up-ward and down-ward branches of frequency chirping. There are also secondary branches in the spectrogram. For a 1D electrostatic wave, Ref. \cite{Berk1997} explains the frequency sweeping at early stages of sweeping as
\begin{equation}
\frac{\delta \omega}{\gamma_l} = \frac{16\sqrt{2} (\gamma_d t)^{1/2}}{3\sqrt(3)\pi^2},
\label{eq:}    
\end{equation}
where $\delta \omega$ represents the frequency shift. Accordingly, Refs. \cite{Boris2010, Nyqvist2012,Nyqvist2013,Hezaveh2017,Hezaveh2020,Hezaveh2021}, provide a theory for long range frequency chirping for unstable eigenmodes in a dissipative background plasma. In these works, the frequency chirping is explained as a self-sustained nonlinear balance between the power extracted from the energetic particles and the power dissipated in the background plasma. To examine that in our self-consistent simulations, we have modified the dissipation coefficients during the frequency chirping. 

We change the dissipation coefficients at $t/t_A=1119.1$ and $1243.4$, from their initial values of \eqref{eq:dis_coef} to $\nu=\eta=6 \cross 10^{-7} v_A R_0$ and $ \nu=\eta=1.2 \cross 10^{-6} v_A R_0$, respectively. We note that this change will not affect the linear evolution of the TAE. \Fref{fig:evolve} shows the resulting amplitude and frequency of the TAE as functions of time that we analyze subsequently. The times at which the damping rate has increased are denoted by vertical dashes in \fref{fig:spec_rise}. A comparison of \cref{fig:evolve_norise,fig:evolve} shows that besides an expected drop in the amplitude of the signals, the rate of frequency chirping has increased in \fref{fig:spec_rise} after increasing the dissipation coefficients. \Fref{fig:evolve} confirms the essential role of dissipation in the chirping mechanism. The above technique of increasing the damping coefficients during the non-linear process of chirping provides a useful probing tool for nonlinear simulations. It can also save computational resources in large-scale simulations. 

\Fref{fig:mode_structure} shows the frequency content at each radial location at four different stages of the wave evolution. The linear mode structure of \fref{fig:spec_rd1} is comparable to the one shown in \fref{fig:radial_profile}. \Fref{fig:spec_rd2} corresponds to the early stages of frequency chirping where the sideband/secondary waves have just formed inside the toroidicity gap. In \cref{fig:spec_rd3,fig:spec_rd4}, the frequencies of the chirping waves deviate further from the initial eigenfrequency towards the tips of the gap which leads to the excitation of continuum waves. Finally, the frequencies of the chirping waves enter the shear \Alfven continuum and exhibit different frequencies at different radial locations as they follow the continuum.            

\begin{figure*}[t!]
\centering
\subfloat[]{
\includegraphics[scale=0.57]{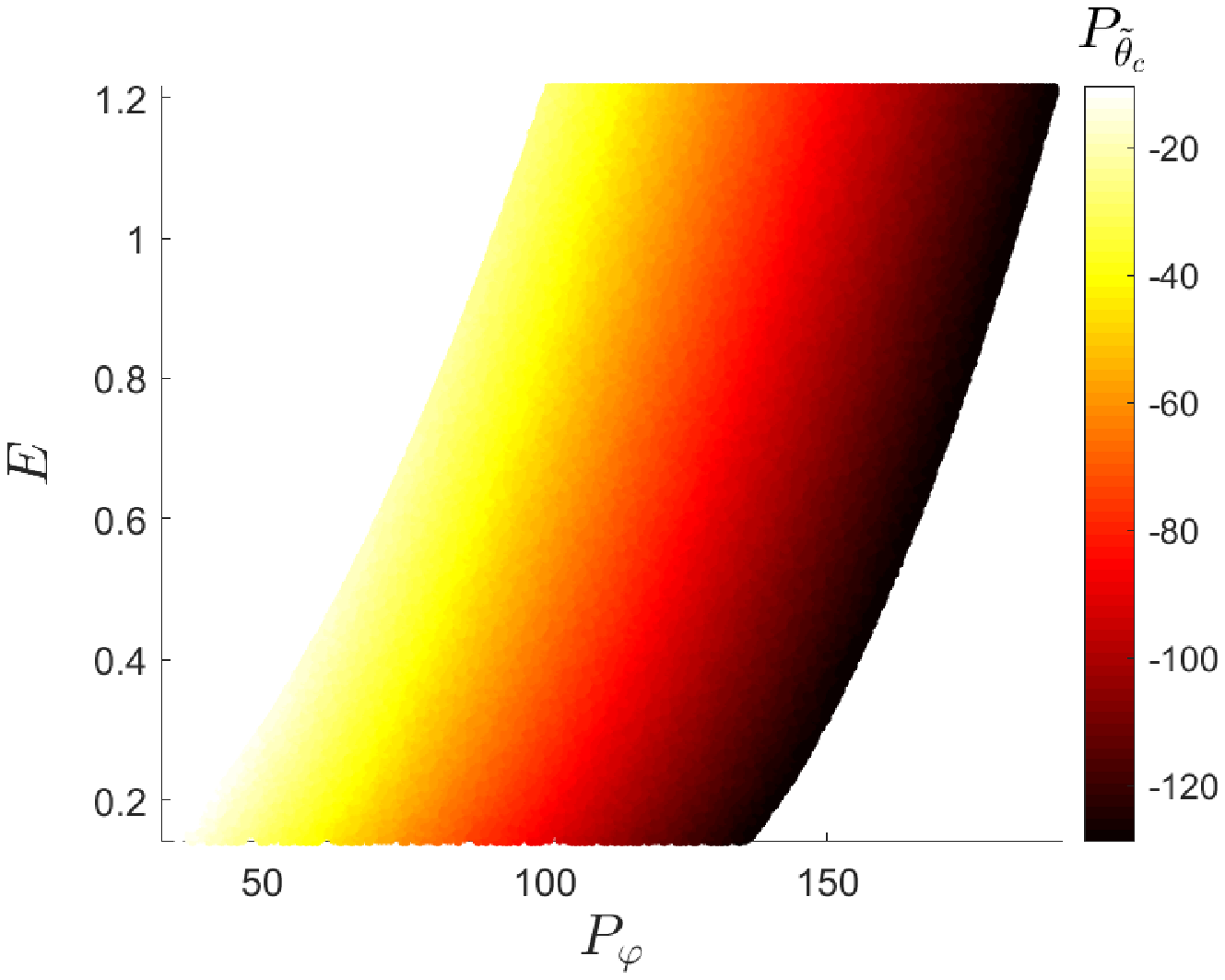}
\label{fig:p_th_E_Pphi}
}
\subfloat[]{
\includegraphics[scale=0.57]{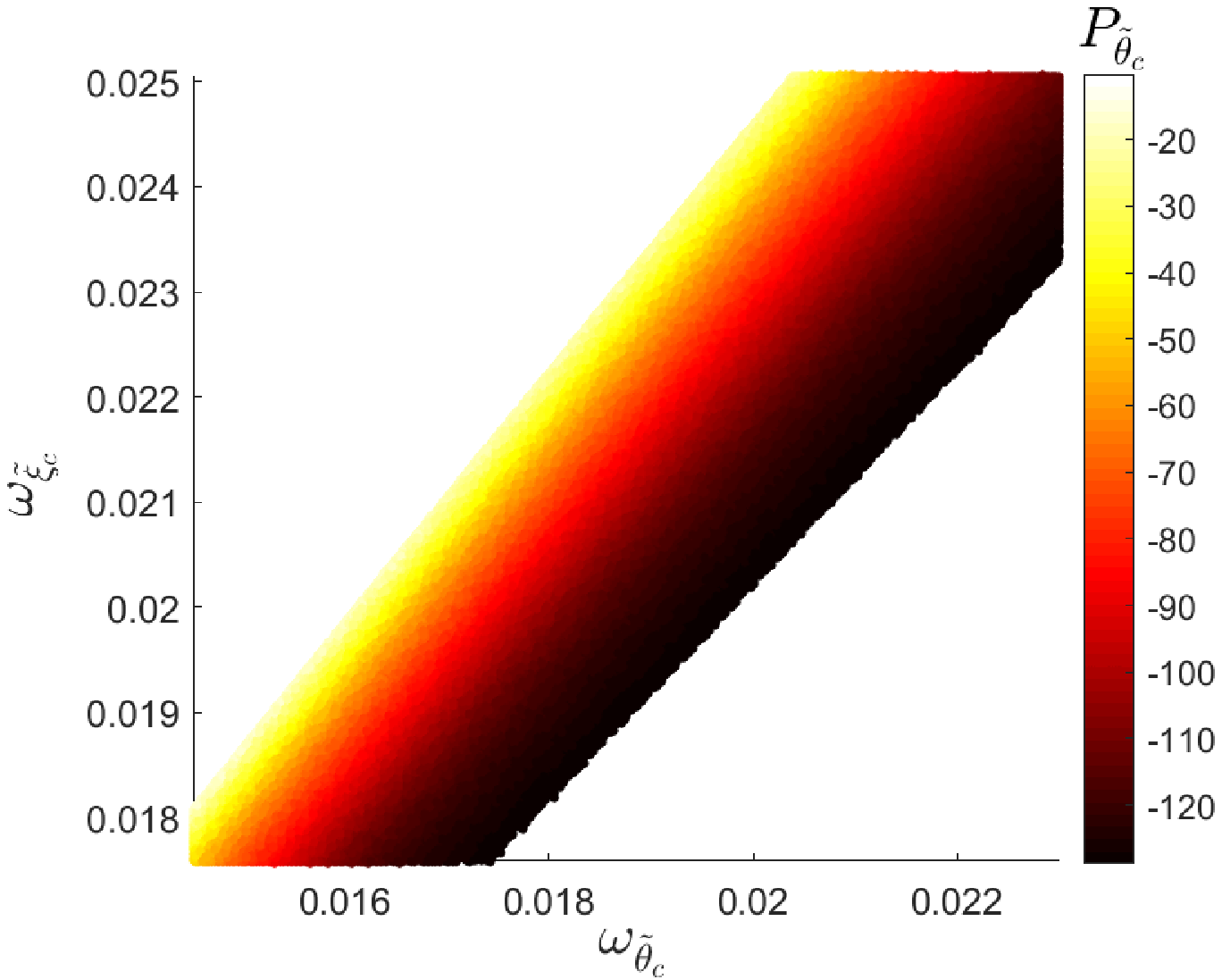}
\label{fig:p_th_omega}
}
\caption{The action corresponding to the poloidal angular momenta for $\mu=0$ in (a) $E$ vs. $P_{\varphi}$ plane and (b) $\omega_{\tilde{\xi}_c}$ VS. $\omega_{\tilde{\theta}_c}$ }.
\label{fig:ps_spc}
\end{figure*}

\subsection{Resonance condition}
In tokamak plasmas, the resonance condition between the particle guiding center motion and a wave with a toroidal mode number $n$ reads \cite{Heidbrink}
\begin{equation}
\omega = n\omega_{\tilde{\xi}_c} + p\omega_{\tilde{\theta}_c}. 
\label{eq:res}
\end{equation}
where $p$ is an integer. In the case of TAE, the mode has two dominant poloidal components of the field ($m$ and $m+1$). These two  components have opposite phase velocities along the magnetic field. Consequently, the strongly co-passing particles resonate at $p=-m$, whereas the strongly  counter-passing particles resonate at $p=-(m+1)$ \cite{Todo1998}. For our modes of interest, we expect the co-passing particle resonance to be at $n=6$ and $p=-6$ in the simulations.

\Fref{fig:ps} shows two images of the perturbed particle distribution function: the color-coded particle weights on the $E-P_{\varphi}$ plane and on the $\omega_{\tilde{\xi}_c}-\omega_{\tilde{\theta}_c}$ plane at the same time.  As expected, the perturbed distribution is strongly localized around the resonance line with $n=6$ and $p=-6$. 

\subsection{Numerical calculation of $P_{\tilde{\theta}}$}
The known frequencies of the unperturbed motion for $\mu=0$ enable calculation of $P_{\tilde{\theta}_c}$ and, thereby, the generating function of the canonical transformation to action-angle variables. To solve this problem, we have used the CVX package \cite{CVX1}. \Fref{fig:p_th_E_Pphi} shows $P_{\tilde{\theta}_c}$ as a  function of $E$ and $P_{\varphi}$. Similarly, the dependence of $P_{\tilde{\theta}_c}$ on the precession frequency $\omega_{\tilde{\xi}_c}$ and the bounce frequency $\omega_{\tilde{\theta}_c}$ is depicted in \fref{fig:p_th_omega}. We observe that for fixed values of $E$, the absolute value of $P_{\tilde{\theta}_c}$ is directly proportional to $P_{\varphi}$. However, a slice of $P_{\varphi}=\text{const}$ demonstrates an inverse relation between the absolute values of $P_{\varphi}$ and $E$.     

\begin{figure}[b!]
    \centering
    \includegraphics[scale=0.59]{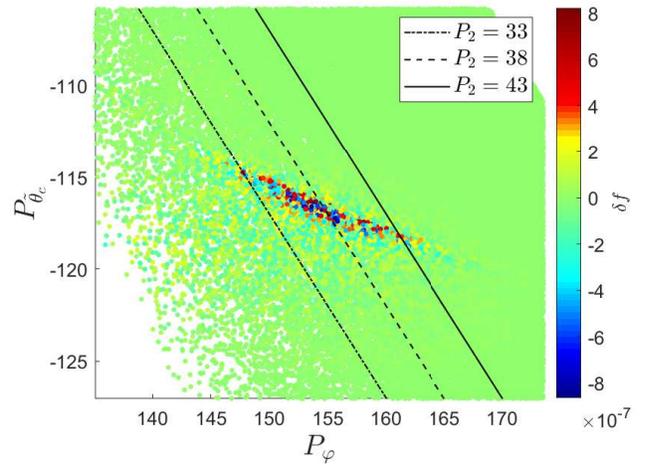}
    \caption{Phase-space dynamics of co-passing EPs on the $P_{\tilde{\theta}}$ vs. $P_{\tilde{\varphi}}$ plane with $\mu=0$ prior to wave saturation. The black lines denote exact constants of motion during the wave evolution.}
    \label{fig:P2}
\end{figure}

At this point, we have introduced all the ingredients to observe/analyse the phase-space dynamics using $(P_1,P_2,P_3)$. \Fref{fig:P2} demonstrates the data of \fref{fig:ps} in the $P_{\tilde{\theta}_c}-P_{\varphi}$ plane. The black lines represent $P_2=\text{const}$ trajectories. Each $P_2=\text{const}$ line corresponds to a sub-layer of the phase-space on which the EPs lie during the evolution of the instability; from the linear phase towards the long range frequency chirping stage. In what follows, we study the detailed dynamics of the resonance in the $P_{1}-\zeta$ plane.

\begin{figure*}
\centering 
\subfloat[$t/t_{A}=205.4$]{\includegraphics[height=0.35\textwidth]{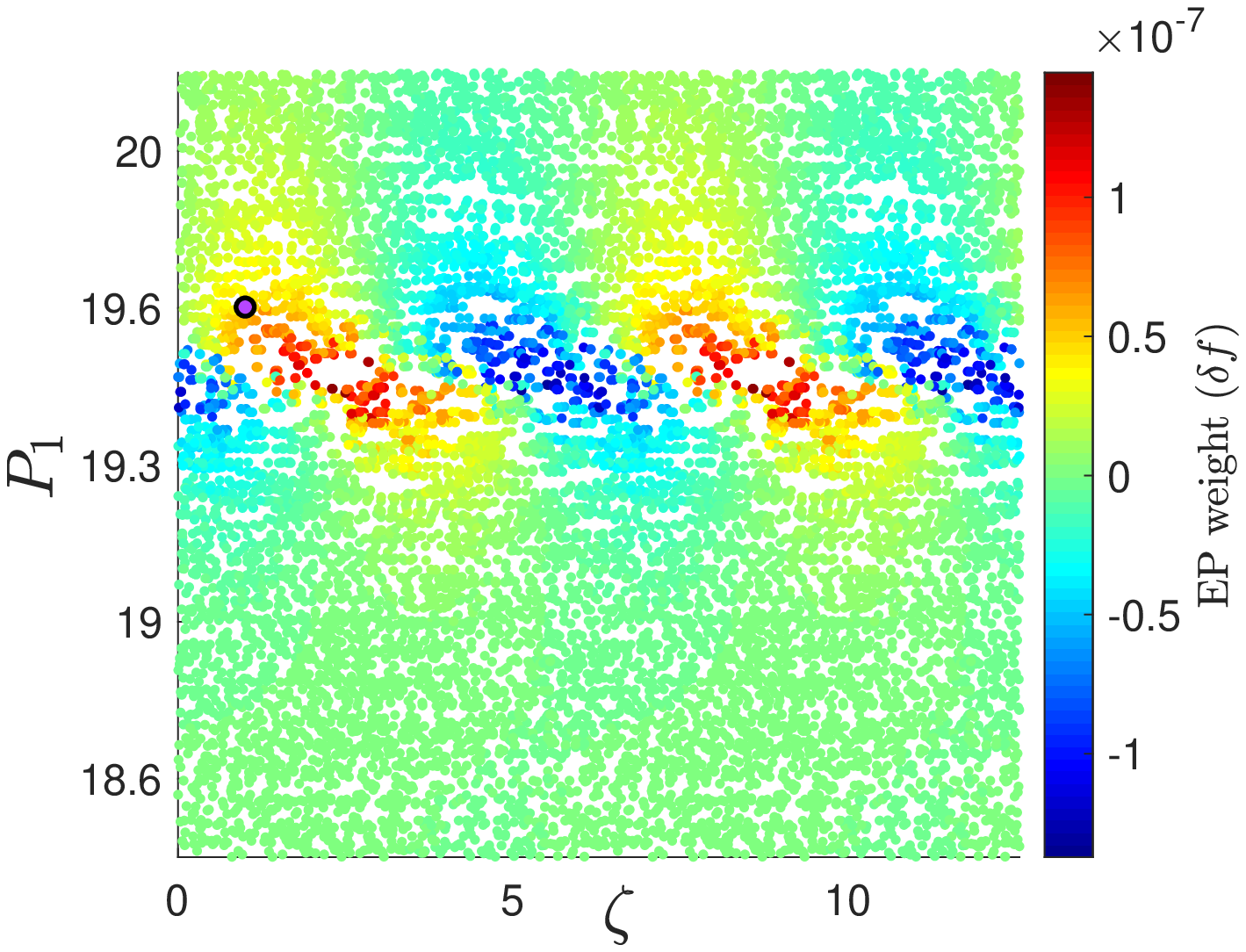} \label{fig:df1}} \qquad
\subfloat[$t/t_{A}=205.4$]{\includegraphics[height=0.35\textwidth]{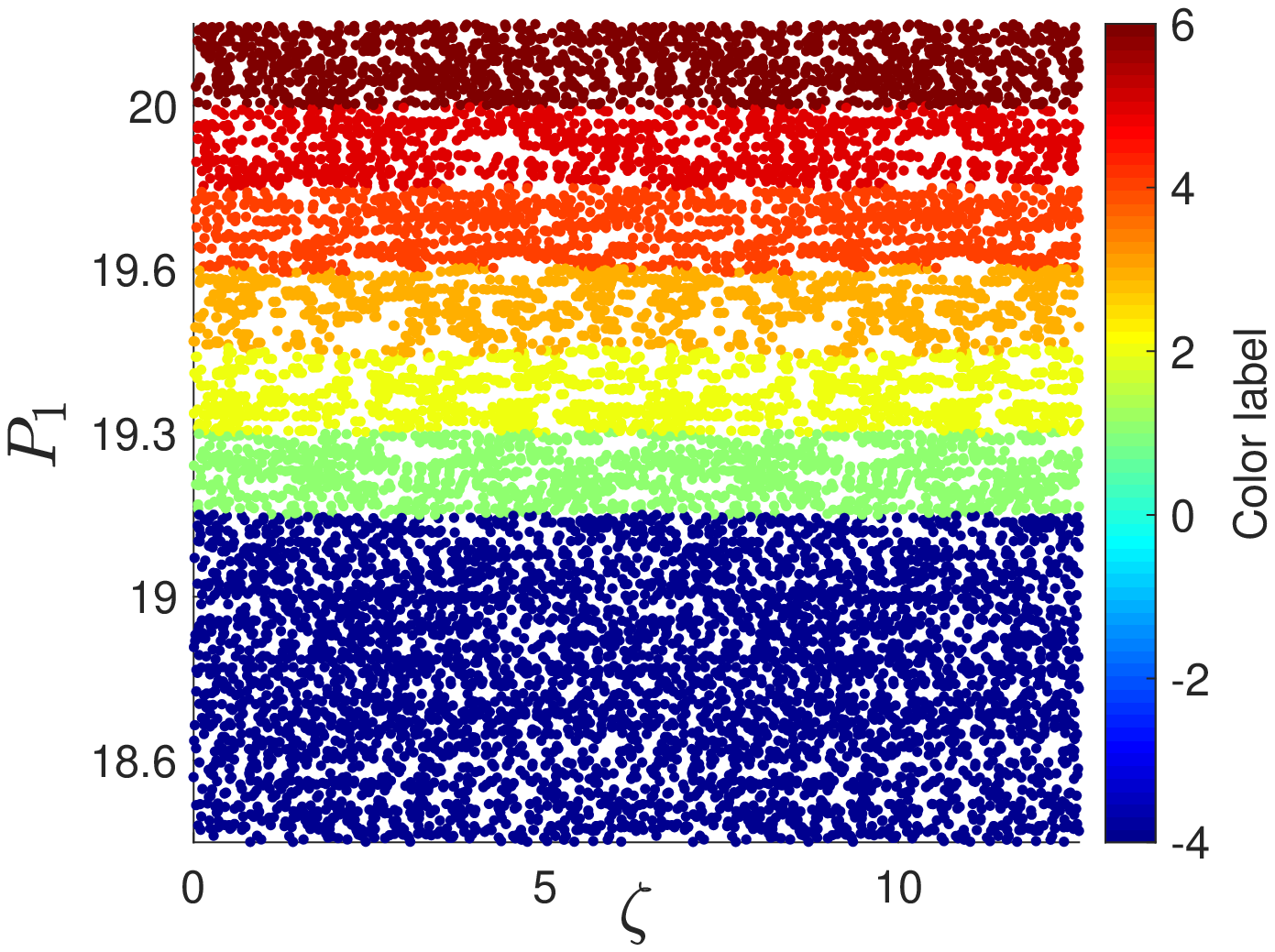}\label{fig:clr1}}  
\rulesep
\subfloat[$t/t_{A}=314.5$]{\includegraphics[height=0.35\textwidth]{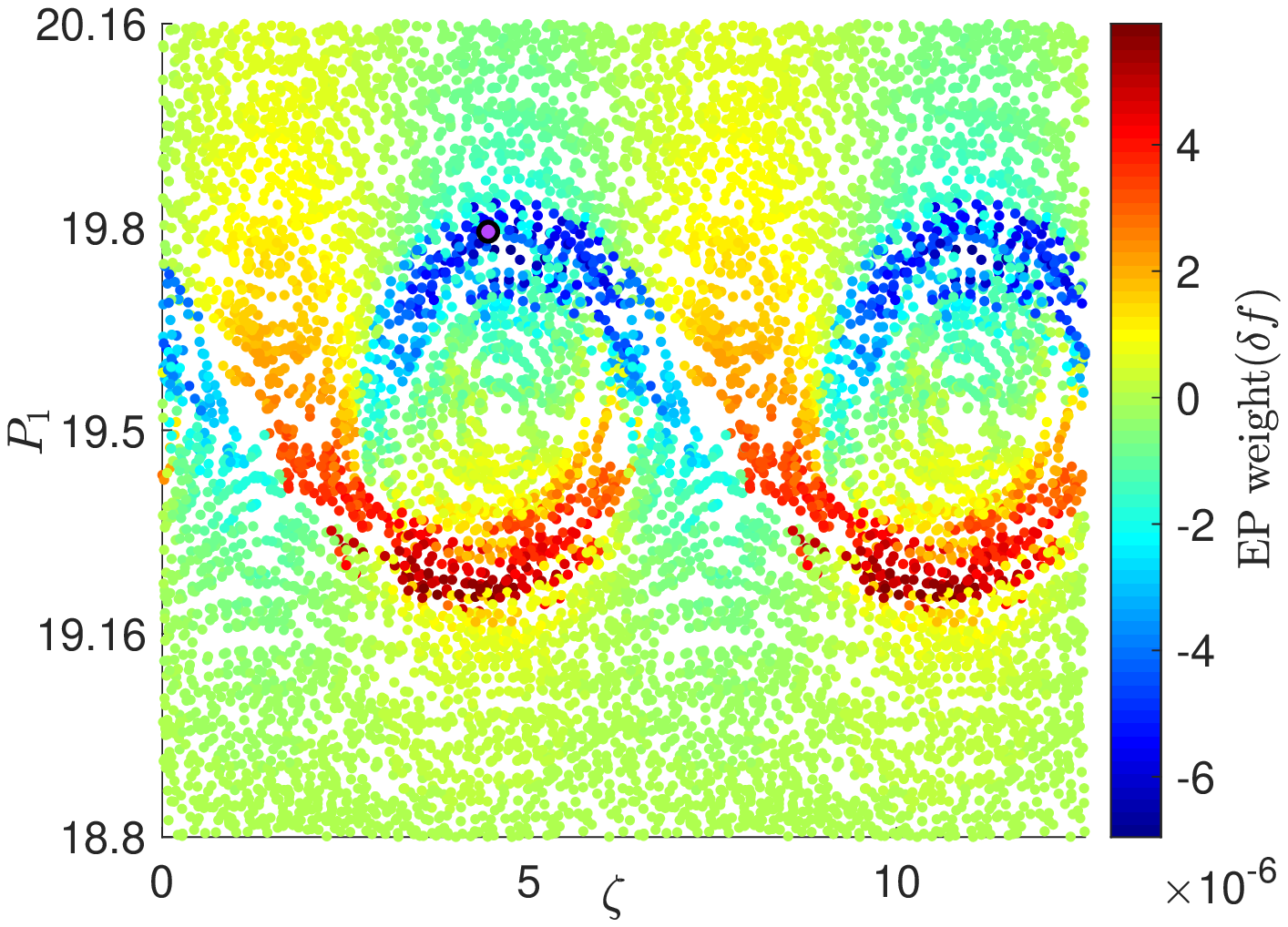}\label{fig:df2}}  \qquad
\subfloat[$t/t_{A}=314.5$]{\includegraphics[height=0.35\textwidth]{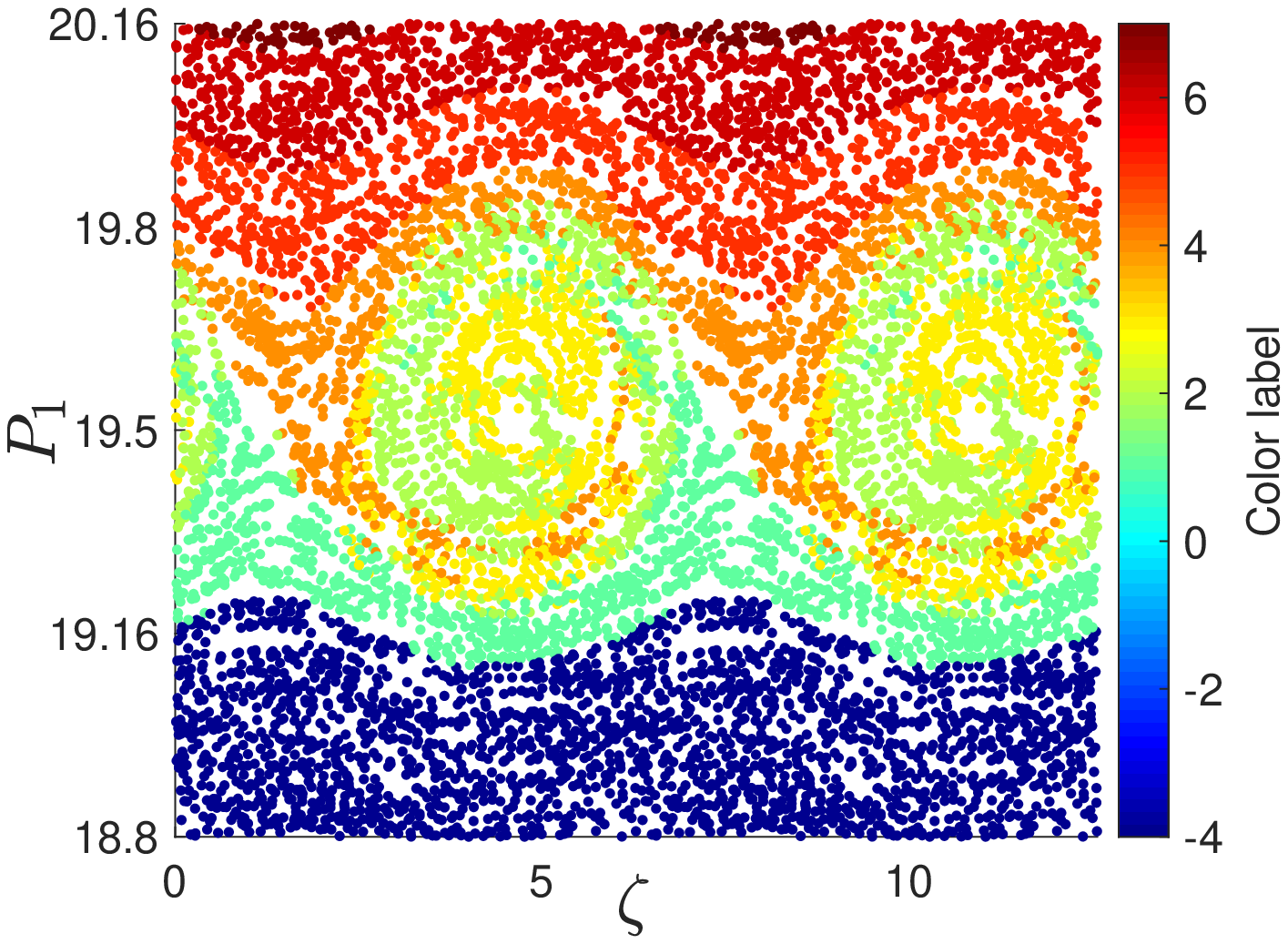}\label{fig:clr2}}  
\rulesep
\subfloat[$t/t_{A}=2477.8$]{\includegraphics[height=0.35\textwidth]{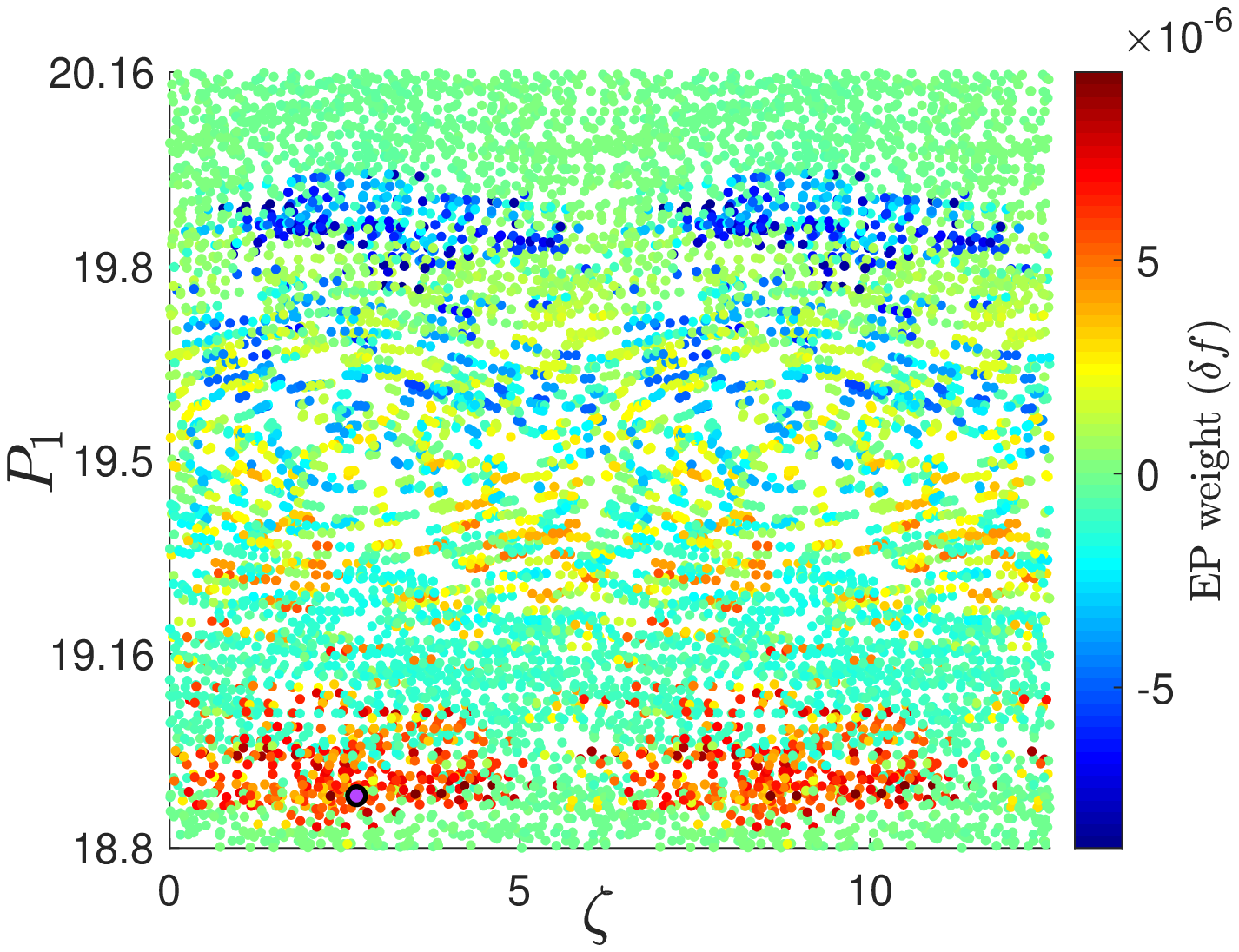}\label{fig:df6}}  \qquad
\subfloat[$t/t_{A}=2477.8$]{\includegraphics[height=0.35\textwidth]{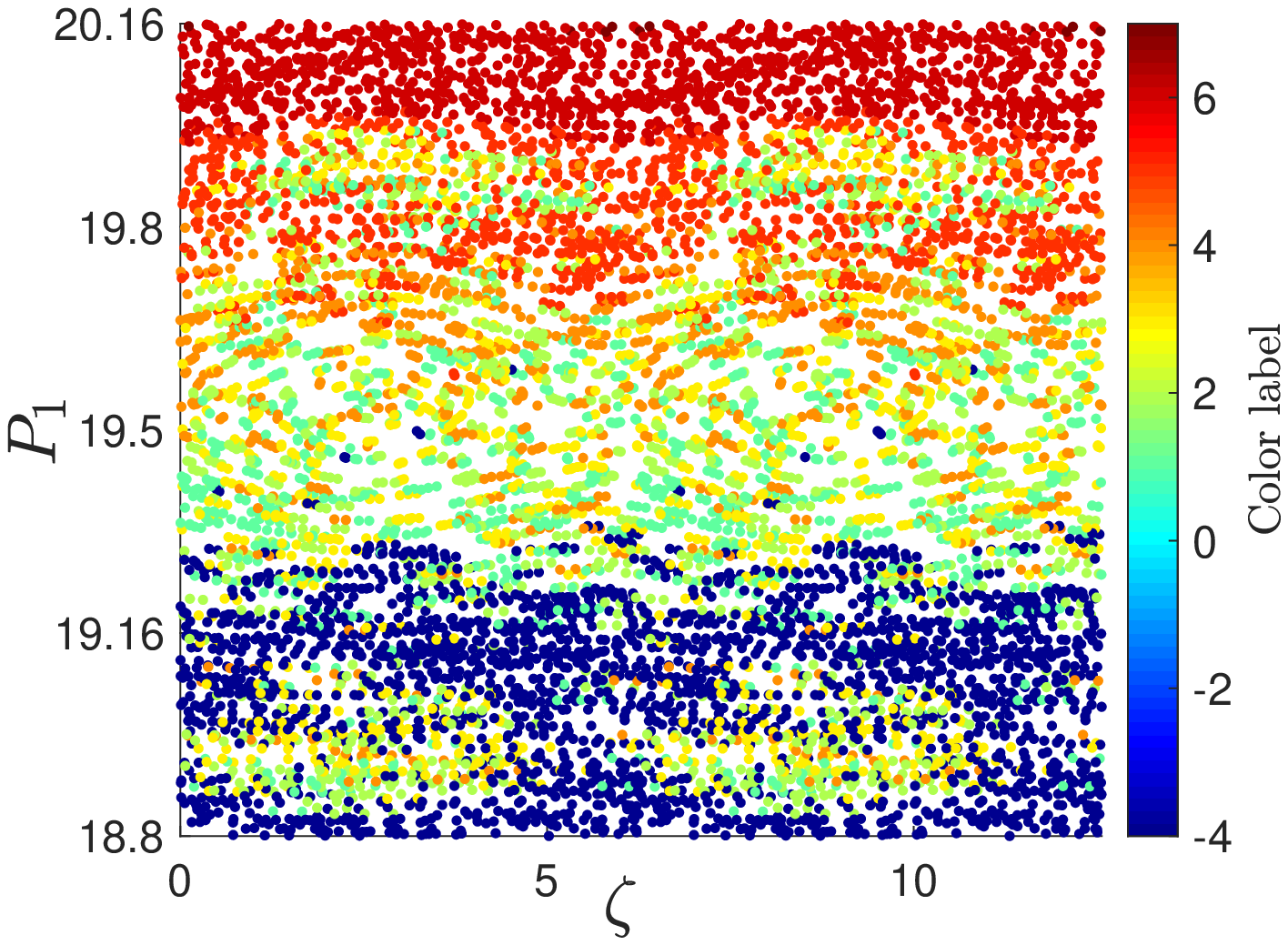}\label{fig:clr6}}  
\rulesep
\caption{A $\mu=0$ and $P_2 \approx 39$ slice of the EPs phase-space as a function of EP weights (panels a,c and e) and EPs color label (panels b,d and f) at different stages of the wave evolution. The purple circle denotes a particle that is convected by the phase-space clump, continued ...}
\end{figure*}

\begin{figure*}
\ContinuedFloat 
\centering 
\subfloat[$t/t_{A}=3224.1$]{\includegraphics[height=0.35\textwidth]{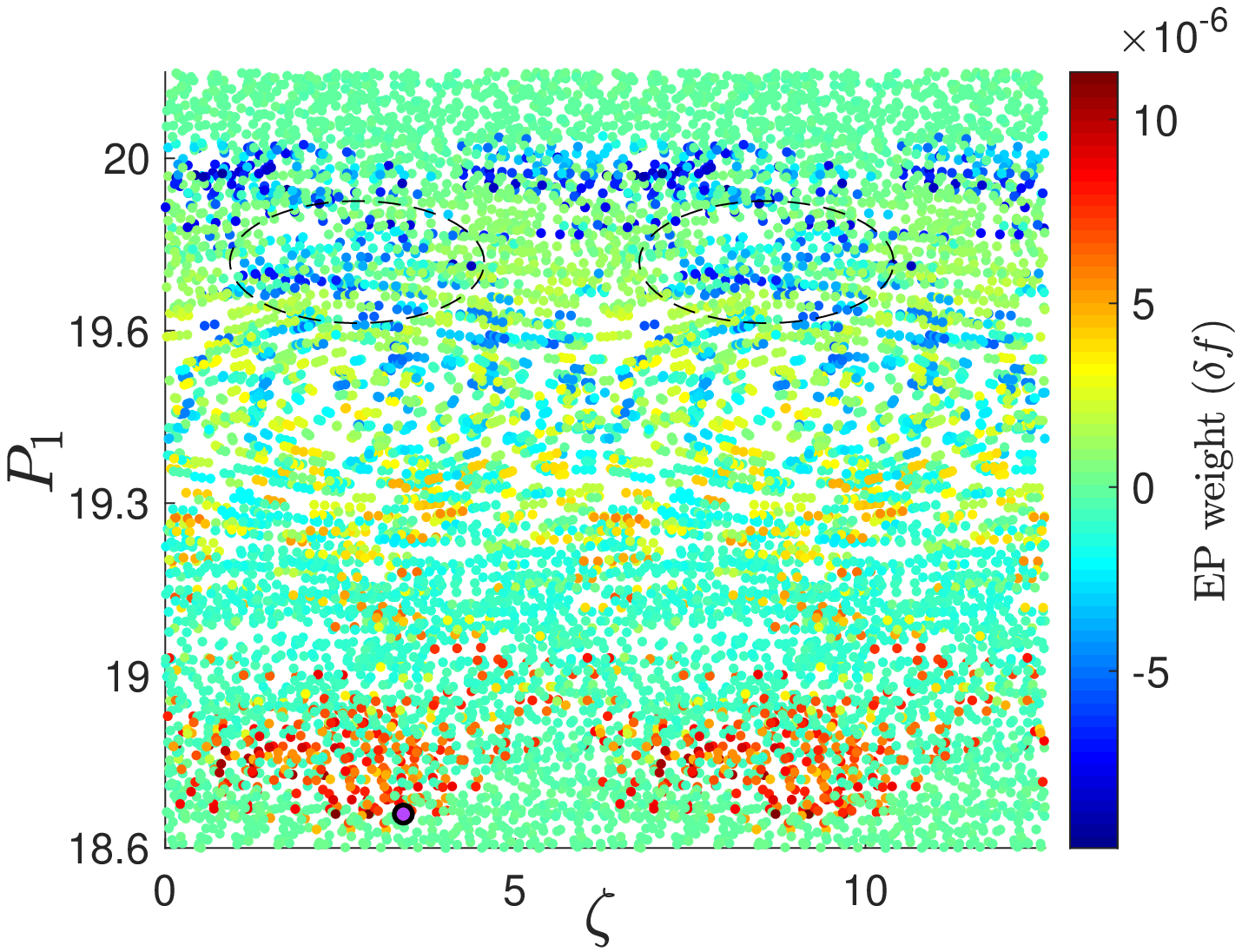}\label{fig:df7}}  \qquad
\subfloat[$t/t_{A}=3224.1$]{\includegraphics[height=0.35\textwidth]{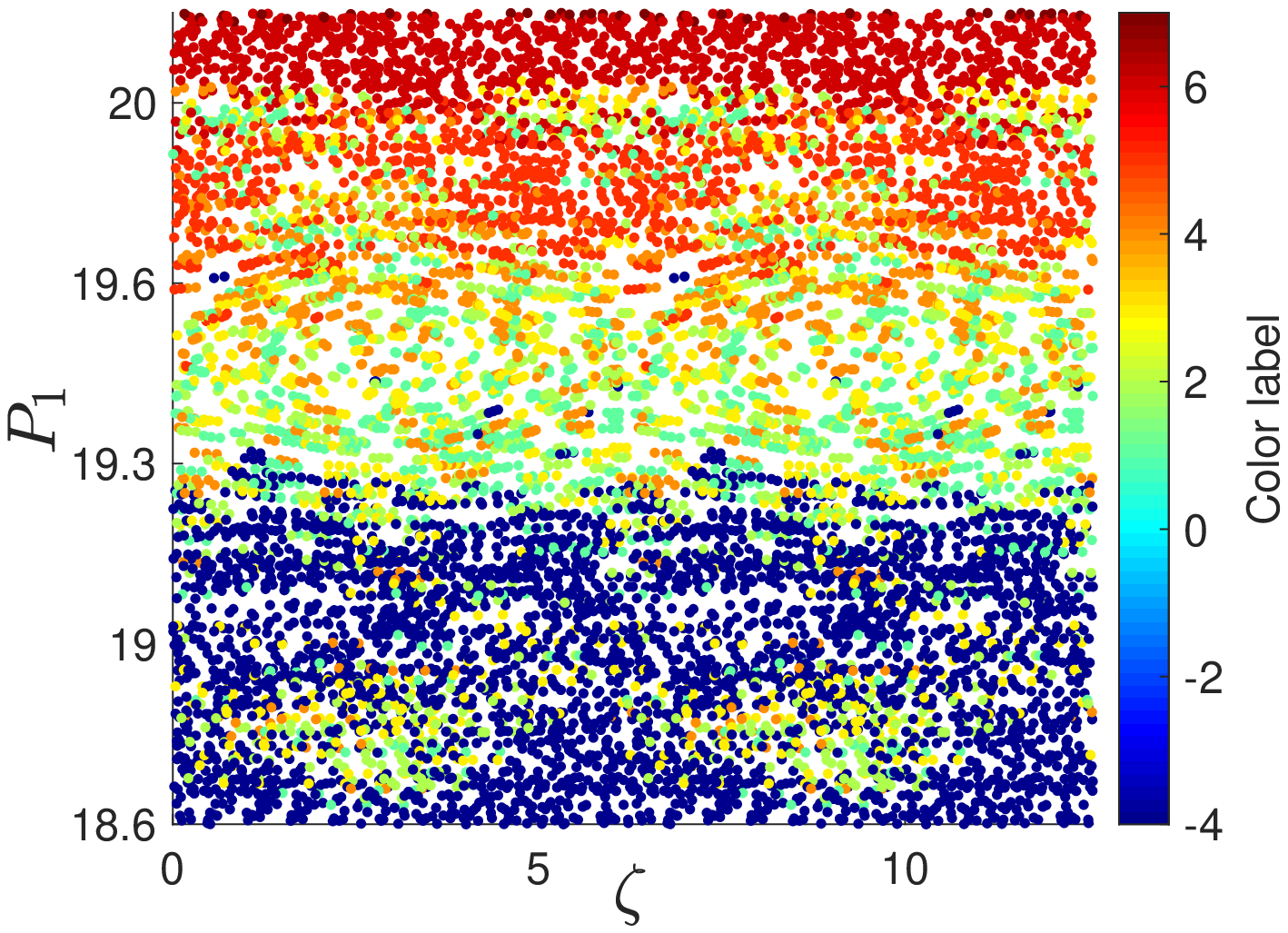}\label{fig:clr7}}
\rulesep
\subfloat[$t/t_{A}=3722.2$]{\includegraphics[height=0.35\textwidth]{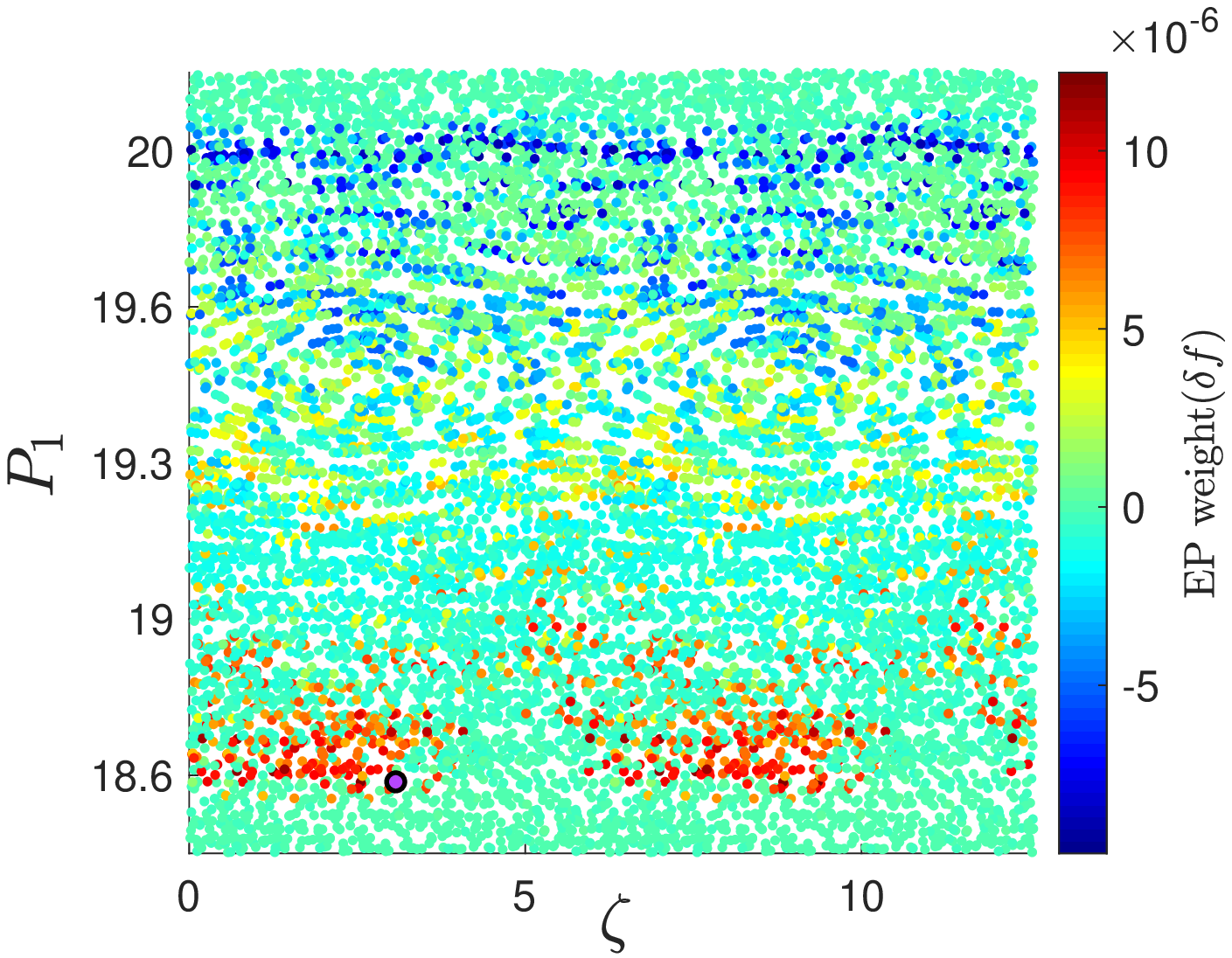}\label{fig:df8}}  \qquad
\subfloat[$t/t_{A}=3722.2$]{\includegraphics[height=0.35\textwidth]{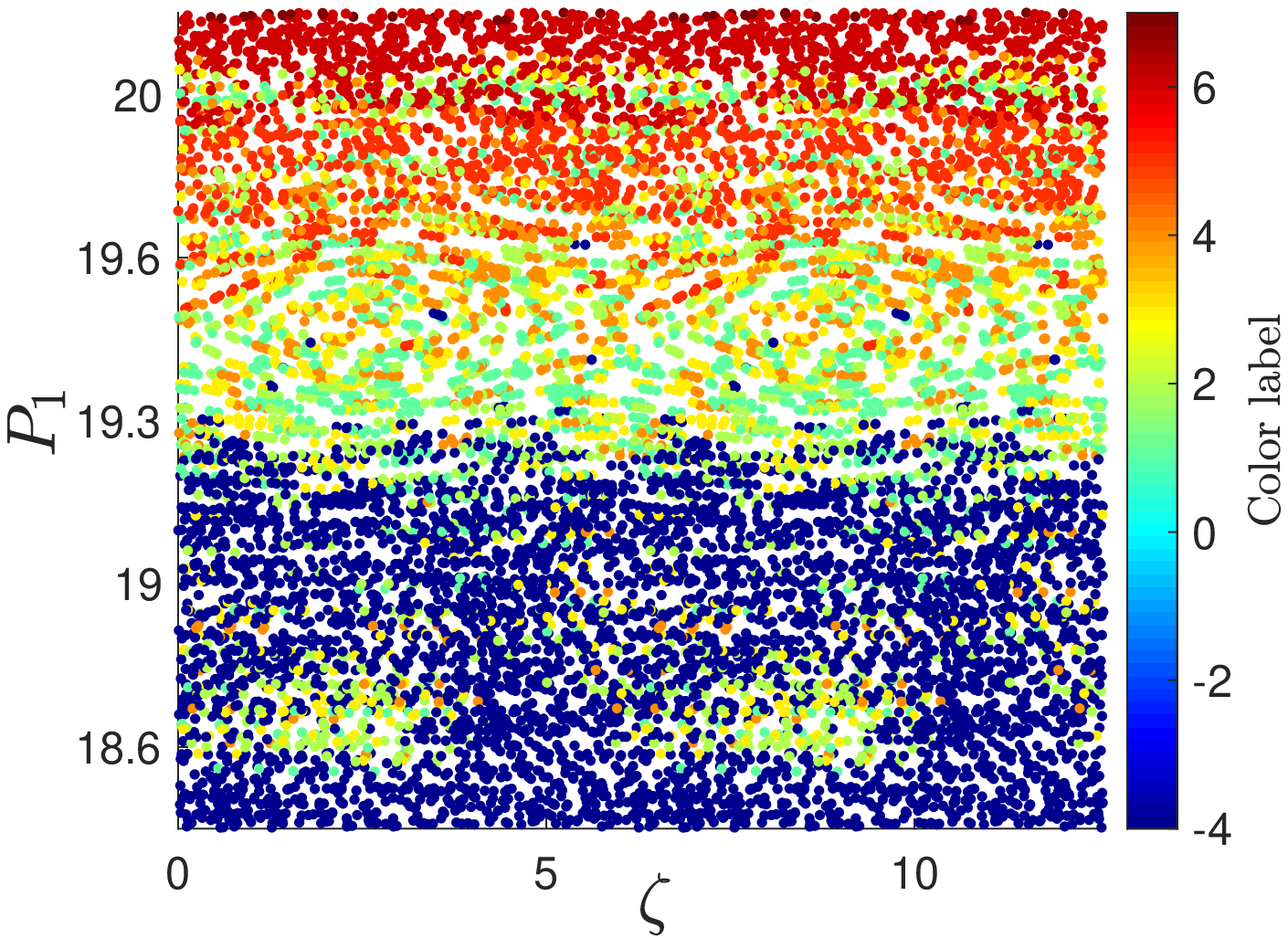}\label{fig:clr8}}
\caption{A $\mu=0$ and $P_2 \approx 39$ slice of the EPs phase-space as a function of EP weights (panels g and i) and EPs color label (panels h and j) at different stages of the wave evolution. The purple circle denotes a particle that is convected by the phase-space clump.}
\label{fig:poincare}
\end{figure*}

\subsection{Convective transport of EPs in phase-space}
In what follows, we analyze a set of particle data recorded at the moments when the particle trajectory crosses the $z=0$ plane with $R>R_0$. The corresponding plots  of $P_1-\zeta$ are essentially \Poincare plots generated for $P_2=\text{const}$ lines in \fref{fig:P2}. We focus on the particles with $P_2=39$. This value is chosen to present EPs with the most perturbed phase-space density (see \fref{fig:P2}). To improve numerical resolution, we record the particle data in the narrow interval $\abs{P_2-39}\leq 0.2$. The aforementioned \Poincare plots are shown in \fref{fig:poincare} at different stages of the TAE evolution. The colors in \cref{fig:df1,fig:df2,fig:df6,fig:df7,fig:df8} represent the perturbed weight/phase-space density of each particle. In the unperturbed state, each EP is assigned a color label according to its corresponding value of $P_1$ (see \fref{fig:clr1}). This label/color is kept the same throughout the simulations. Using this label, we produce a set of snapshots of the phase-space i.e. \cref{fig:clr1,fig:clr2,fig:clr6,fig:clr7,fig:clr8} where the color bar denotes the particle label. In addition,  we use this label to see that the particles resonant with the chirping waves are not being locally perturbed and then left behind. Rather, the BGK-type chirping wave carries the EPs in a moving phase-space bucket in a convective way. 

\begin{figure}
    \centering
    \includegraphics[scale=0.59]{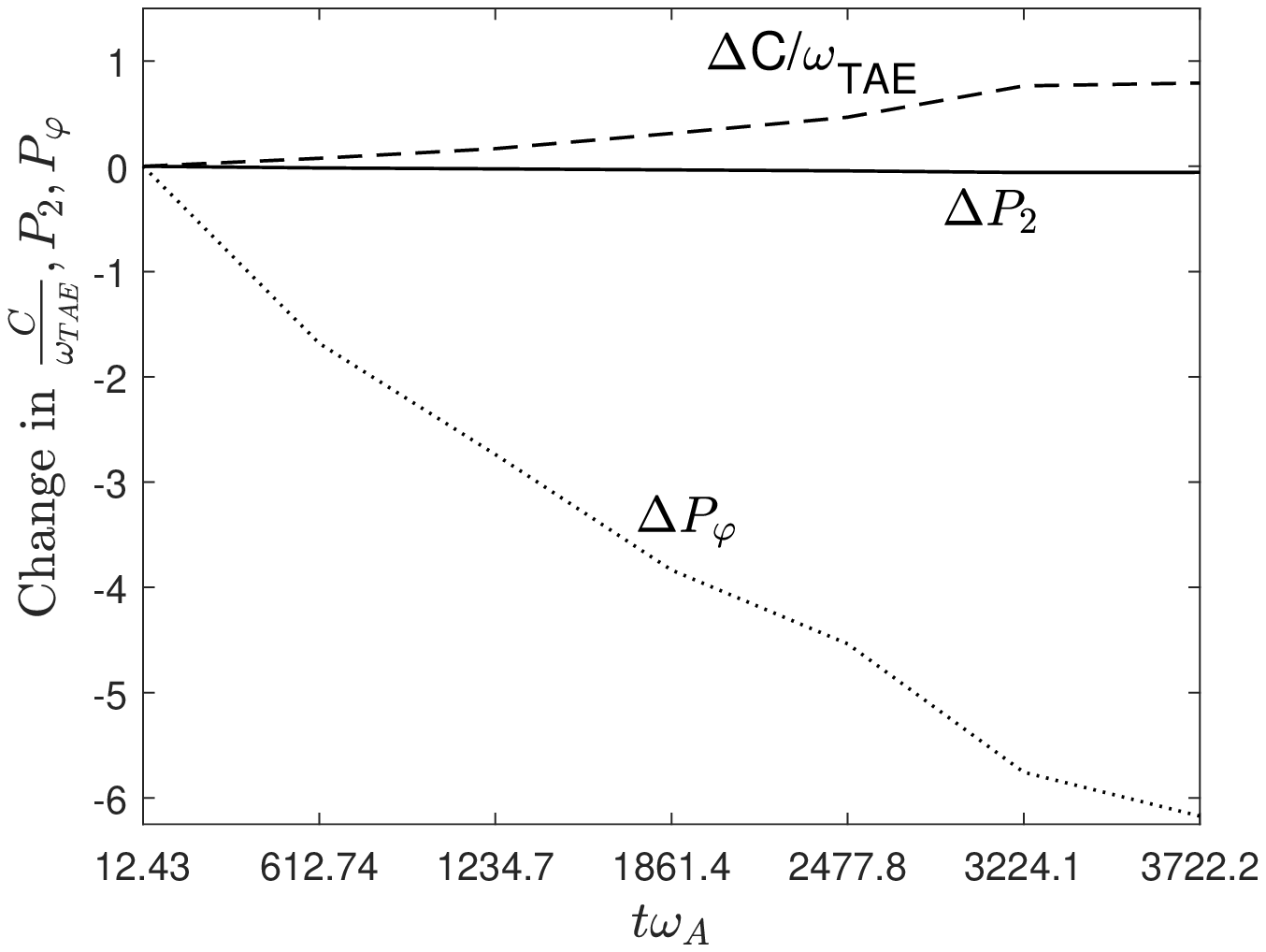}
    \caption{Variation of $\frac{C}{\omega_{\text{TAE}}}$, $P_2$ and $P_{\varphi}$ at various stages of the TAE evolution.}
    \label{fig:change}
\end{figure}

\Cref{fig:df1,fig:clr1} correspond to the linear stages of the TAE excitation i.e. $t/t_{A}=205.4$. \Cref{fig:df2,fig:clr2} demonstrate the coarse graining of the distribution function in phase-space just before the non-linear saturation of the TAE. \Cref{fig:df6,fig:clr6} demonstrate the phase-space dynamics during the frequency chirping of the wave at $t/t_A=2477.8$. At this point, the up-chirping and down-chirping waves have experienced a frequency sweep of $13.55\%$ and $11.33\%$, respectively. We observe the holes (blue) and clumps (red), corresponding to the down-chirping and up-chirping waves, respectively. They form at either side of the flattened region and move in the phase-space of EPs as the frequencies chirp. The rest of the panels correspond to further evolution of the frequencies. It is worth mentioning that the dashed ovals in \fref{fig:df7} mark the detachment of a second set of phase-space holes. We attribute these structures to the second branch of down-chirping waves illustrated in \fref{fig:spec_rise}. 

Since EPs remain on the same sub-layer of the phase-space, on which $P_2$ is a constant of motion, the constructed phase-space plots ascertain the mechanism under which the phase-space density is being perturbed. As \cref{fig:clr6,fig:clr7,fig:clr8} clearly demonstrate, the phase-space islands act like buckets that carry particles in phase-space and lead to radial convection of the EPs. Conservation of the generalised momentum $P_2$, given by \eqref{eq:CT1}, is the key part of this understanding. Although the constancy of $P_2$ is evident in phase-space plots of \fref{fig:poincare}, we investigate the value of $P_2$ as a function of time for an EP which is transported by the up-chirping wave. This particle is denoted in \cref{fig:df1,fig:df2,fig:df6,fig:df7,fig:df8} by a purple circle. Simultaneously, we calculate the value of $C$, introduced in section \ref{sec:intro}, for the same EP. This comparison is depicted in \fref{fig:change} where the value of $C$, unlike $P_2$, changes as the mode frequency begins to chirp. It is worth noting that $P_2$ is comparable to $\frac{C}{\omega_{\text{TAE}}}$ in terms of units. Hence, slices of $C=\text{const}$ do not represent the most appropriate sub-layers of the phase-space to study/observe the dynamics during the long range frequency chirping.

\section{Summary}
\label{sec:summary}

We have refined the formalism for the phase-space analysis of the chirping modes driven by resonant energetic particles in a tokamak. As an application of this refinement, we analyze the results of self-consistent simulations performed with the MEGA code (an initial value problem solver in a hybrid MHD-kinetic model). The initial perturbation under study is a shear \Alfven eigenmode in the toroidicity-induced gap of the \Alfven continuum (TAE). The initial population of the energetic particles has an isotropic slowing down distribution. The EPs current provides a linear growth drive of $\gamma_l/ \omega_{\text{TAE}}=2.64 \% $ to the mode in the presence of background dissipation at a rate of $\gamma_d/\gamma_l=0.44$.  

Subsequent to the non-linear saturation of the eigenmode, the sideband (secondary) oscillations appear inside the toroidicity gap. These modes evolve into chirping waves. In this case, we observe both up-ward and down-ward trends as the frequency chirps. We demonstrate that the rate of frequency sweeping increases with the damping rate of the eigenmode. As the chirping waves enter the shear \Alfven continuum, the radial structure of the perturbation experiences different frequencies at different radii. This is consistent with the theoretical model of Ref. \cite{Wang2018}. 

Investigation of the energetic particle dynamics reveals that these particles lie on the same sub-layer of the phase-space throughout the simulations. Contingent on the formation and evolution of the chirping waves, phase-space islands form and evolve adiabatically. This means that the same particles are carried inside the coherent phase-space islands providing a convective or bucket transport in phase-space. Once formed in the gap, the phase-space holes and clumps survive even in the shear \Alfven continuum.    

\section*{Acknowledgments}
This work was funded by the Australian Research Council through Grant No. DP140100790 and supported by the U.S. Department of Energy Contract No. DEFG02–04ER54742. This research was undertaken with the assistance of resources and services from the National Computational Infrastructure (NCI), which is supported by the Australian Government. The National Institutes of Natural Sciences (NINS) and National Institute for Fusion Sciences (NIFS) have supported two internships of the first author at NIFS, Japan in 2018 and 2019. The first author is very thankful to Prof. Yasushi Todo, Prof. Masayuki Yokoyama, Prof. Hao Wang, Dr. Malik Idouakass and Dr. Jialei Wang for their kind hospitality and fruitful discussions during his stay at NIFS, Japan.

\section*{References} 

\bibliography{references}
\bibliographystyle{iopart-num}

\clearpage

\end{document}